\documentclass[11pt]{article}

\usepackage{amssymb,slashed,graphicx,color,epsf}
\usepackage{blkarray}
\usepackage{amsmath}
\usepackage[nosort]{cite}
\usepackage{cancel}
\usepackage{pdflscape}
\usepackage{rotating}
\usepackage{float}
\usepackage{xcolor,colortbl}

\usepackage{tcolorbox} 
\tcbset{
  colback=white    
}

\def\tr{{\rm tr}}

\def\divQ{\mathrm{div}Q}
\def\divQt{\mathrm{div}\tilde Q}
\def\trdivQ{\mathrm{trdiv}Q}
\def\trdivQt{\mathrm{trdiv}\tilde Q}
\def\nn{\nonumber}

\def\be{\begin{equation}}
\def\ee{\end{equation}}
\def\ben{\begin{displaymath}}
\def\een{\end{displaymath}}
\def\bea{\begin{eqnarray}}
\def\eea{\end{eqnarray}}

\makeatletter \@addtoreset{equation}{section} \makeatother

\def\cL{{\cal L}}

\def\cP{{\cal P}}

\def\alsix{\alpha}

\newcommand{\w}[1]{\\[0.#1cm]}

\def\eq#1{(\ref{#1})}

\newcommand{\hoch}[1]{$\, ^{#1}$}

\def\tr{{\rm tr} }

\def\tQ{{\widetilde Q}}

\newcommand{\tred}{\textcolor{red}}

\def\m{\mu}
\def\n{\nu}
\def\r{\rho}
\def\s{\sigma}

\newcommand{\tamphys}{\it George and Cynthia Woods Mitchell  Institute
for Fundamental Physics and Astronomy,\\
Texas A\&M University, College Station, TX 77843, USA}

\newcommand{\auth}{\large
Roberto Percacci\hoch{\sharp}
and Ergin Sezgin\hoch{\dagger}
}

\begin{document}

\begin{flushright}\small
MI-HET-863   \\
\end{flushright}

\vspace{25pt}

\begin{center}

{\Large \bf Massive spin 3 and Metric-Affine Gravity}

\vspace{25pt}
\auth

\vspace{20pt}
\hoch{\sharp}{\it  SISSA, via Bonomea 265, Trieste, Italy and INFN, Sezione di Trieste.}

\vspace{10pt}
\hoch{\dagger}{\tamphys}

\vspace{40pt}

\underline{ABSTRACT}
\end{center}

Symmetric Metric-Affine Gravity is a theory of gravity with an independent non-metric connection,
and zero torsion. It can be thought of as ordinary metric gravity coupled to a rank--3 tensor $Q$, symmetric in one pair of indices. This field carries a spin-3 degree of freedom. We find that, in contrast to the totally symmetric case, it is possible to arrange the parameters in the Lagrangian, so that, at the linearized level, the spin-3 state, either alone or in combination with a spin-0 state, has a healthy propagation.

\vspace{15pt}

\thispagestyle{empty}

\pagebreak


\tableofcontents

\newpage

\section{Introduction}

Particle states can be viewed as irreducible representations (irreps) of the Poincar\'e group
and in particle physics we are accustomed to using one field for each particle species. In particular, every boson of spin $s$ is associated to a totally symmetric tensor of rank $s$. 
\footnote{ This is the situation for fields with parity $P=(-1)^s$. For fields with opposite parity, see Appendix \ref{standardpropagators}.}
This, however, is a reducible representation, and only the highest dimensional irrep is used. If the free Lagrangian is chosen so that the spin $s$ components propagate correctly, the other, lower spin, components will typically present some pathologies. For example, the Proca Lagrangian has a specific form such that the spin zero (longitudinal) component of the vector does not propagate, because if it did it would be a ghost. A similar situation arises  in the linearized Einstein Lagrangian.
In higher spin theories, however, the number of irreps is larger and it becomes possible to find Lagrangians for a single field that correctly propagate more than one type of particle.
One context where this happens is Metric-Affine Gravity (MAG), a vast class of generalized theories of gravity where the connection is independent of the metric~\cite{Hehl:1994ue,Blagojevic:2013xpa}.
The connection is not a tensor and it is more convenient to use as an independent
variable the difference between the independent connection and the
Levi-Civita connection, which is a tensor.
One can then recast MAG as a normal metric theory of gravity coupled to a rank three tensor.
When linearized around flat space, this field carries in general several irreps, including one with spin 3.
This is therefore a natural context in which a higher spin field arises in a theory of gravity.
It has been known for a long time that there are examples of MAGs where the connection correctly describes the propagation of more than one type of particle
\cite{Sezgin:1979zf,Lin:2018awc,Aoki:2019snr}.
Motivated by these examples, we shall seek free MAG
Lagrangians that,
at the linearized level, correctly describe
the propagation of a massive spin-3 particle and possibly some other lower spin state, in addition to the graviton. 

An independent motivation for this work comes from the theory of higher spin fields. 
It is well known since the work of Singh and Hagen that the description of a massive higher spin field involves, in addition to the totally symmetric tensor, also auxiliary fields \cite{Singh:1974qz}. 
In particular, in the case of spin 3, one auxiliary scalar is used. 
One may wonder whether there are ways of describing the massive spin 3 particle without introducing additional fields,
or in other words whether some of the lower spin states carried
by the rank-3 tensor can play the role of the auxiliary fields.
Whereas normally one demands that the rank-$s$ tensor describe
the propagation of a spin-$s$ {\it only},
we can relax this condition and allow for the propagation of
other lower spin states.
We shall see that for $s=3$, as long as we use a totally symmetric tensor, 
this is not enough: 
there is no free Lagrangian for a totally symmetric rank-3 field
that describes the propagation of a massive spin-3 particle and any combination
of lower spin states, avoiding ghosts and tachyons.

At this point, partly motivated by MAG, we will relax the condition of total symmetry and demand that the field be symmetric in two indices only.
(In the geometrical language of MAG this corresponds to allowing the
connection to carry nonmetricity, but to have zero torsion.)
In this case, we will see that it is possible to find (many) Lagrangians
that propagate either only a massive spin-3 particle or a massive spin-3 and a massive spin-0 particle (other combinations may also be possible, but we have not
systematically explored all possibilities). 
In a sense, in this way we did not have to introduce auxiliary fields.
However, the symmetric field can be decomposed algebraically and uniquely
in a totally symmetric part and a hook symmetric part
(symmetric in $a,b$ but such that $\Phi_{(abc)}=0$).
For those Lagrangians that propagate a spin-3 particle only, one could view the hook symmetric part as being an auxiliary field.  
In our models we do not have lower rank fields, except possibly for the rank-2 graviton field $h_{\mu\nu}$, that could mix with some of the spin $2^+$ components of the three-tensor.
This field has to satisfy the linearized Einstein equations. It lives a life of its own and will be ignored in most of this work.

We conclude with some remarks about the earlier literature on spin 3 and MAG. Baekler, Boulanger and Hehl \cite{Baekler:2006vw} found a choice of parameters in a  gauge theory of gravity, such that at linearized level it describes the propagation of a spin $3^-$ state.
We comment more extensively on their work in Section \ref{discussion}.
For related results using the spin projector formalism see \cite{Baldazzi:2021kaf,Mikura:2024mji,Barker:2025xzd,Barker:2025rzd}.
All these papers discuss the embedding of the (massless) Fronsdal Lagrangian in the linearized MAG Lagrangian.
Their limitation is that it seems very unlikely that the tensorial gauge symmetry of the Fronsdal Lagrangian can be extended to a fully interacting theory, as has been reviewed in \cite{Bekaert:2010hw}. For this reason, it seems more promising to look for a massive spin $3^-$ field, that does not have such a symmetry.
Using the spin projector formalism, Marzo  found a Lagrangian for massive spin $3^-$ and spin $1^-$ \cite{Marzo:2021esg}.
His approach differs from ours in that he used a vector and a scalar as auxiliary fields. The aim of this paper is to not introduce
additional lower rank fields; instead, in a sense, we use the hook symmetric components of nonmetricity as auxiliary fields. This seems more natural in MAG, where nonmetricity is in general a three-index field, symmetric in two indices only.

\smallskip

The paper is organized as follows. In Section 2 we briefly describe the spin projector toolbox. 
In Section 3 we treat the case of a totally symmetric rank-3 tensor field and in Section 4 we generalize to the case of a rank-3 tensor field that is symmetric in two indices only. The relevance of these results to MAG and aspects of interactions is discussed in Section 5.

\section{The spin projector toolbox}

In determining the spectrum of the theory and analysing the ghost and tachyon freedom conditions, it is convenient to use the spin projector formalism.
Assume that the dynamical field is $\Phi_A$, where $A$ stands for a set of Lorentz indices.
The field can be decomposed in irreducible representation of the rotation group $O(3)$, that are characterized by their spin $J$ and parity $\cP$.
There may be several representations with the same spin and parity, in which case they are distinguished by an index $i$.
For each such representation with label $i$ there is a projector
$\big(P(J^\cP)_{ii}\big)_A{}^B$
and for each pair of representations with the same $J^\cP$ and labels $i$, $j$ there is an intertwiner
$\big(P(J^\cP)_{ij}\big)_A{}^B$.
The full set of the $\big(P_{ij}(J^\cP)\big)_A{}^B$ is orthonormal and complete:
\be
P_{ij}(J^\cP)\cdot P_{kl}(I^{\cal Q})=
\delta_{IJ}\,\delta_{\cal PQ}\,\delta_{jk}\,\ 
P_{il} (J^\cP)\ ,\qquad 
\sum_{J,\cP,i} P_{ii}(J^\cP)={\bf 1}\ ,
\ee
where we have suppressed the $A,B$ indices of the projection operators and their momentum labels.

The quadratic part of the action, in momentum space, can be written in the general form
\be
S^{(2)}=\int \frac{d^4 q}{(2\pi)^4}\ \left[\frac12 
\sum_{J,\cP,i,j}\Phi_A(-q)\cdot
a_{ij}(J^\cP,q)\,P_{ij}^{AB}(J^\cP,q)\cdot \Phi_B(q) + {\cal J}^A(-q)\cdot \Phi_A(q)\right]\ ,
\label{adam}
\ee
where, in the case of interest to us, $\Phi_A=Q_{\lambda\mu\nu}$
are the fields and 
${\cal J}^A=\tau^{\lambda\mu\nu}$ are the sources they couple to.
Solving the classical equations of motion, one obtains for the
propagator, saturated with external sources,
\bea
\Pi &=& -\frac12\int \frac{d^4 q}{(2\pi)^4}\  {\cal J}^A(-q)\cdot 
a^{-1}_{ij}\,P_{ijAB}(J^\cP,q)\cdot {\cal J}^B(q)
 \ .
\label{eve}
\eea
Strictly speaking the above formula is only valid when the matrices $a_{ij}$
are invertible, which is the case in the absence of gauge invariances.
When gauge invariances are present, $a^{-1}_{ij}$ has to be replaced by
$b^{-1}_{ij}$, where $b_{ij}$ is a submatrix of $a_{ij}$ of maximal rank.
We will not have to consider this complication.

The information about the propagating degrees of freedom is contained in the 
determinants of the matrices $a_{ij}(J^\cP)$.
These are polynomials in $q^2$ that can be written in the form
\be
C(q^2+\mu_1^2)\ldots (q^2+\mu_s^2)
\ee
where $C$, $\mu_1$,\ldots,$\mu_s$ are real constants.
Absence of tachyons will be guaranteed provided $\mu_i^2>0$ for all $i$.

The saturated propagator $\Pi$ will have poles at $q^2=-\mu_i^2$,
and the absence of ghosts demands that residues at the poles be positive.
There is a further sign due to the fact that the projectors in (\ref{eve})
have lowered Lorentz indices compared to those in (\ref{adam}).
Since the momentum in the projectors is timelike,
in our signature $-+++$ there will be a factor $(-1)^{n_L(J,P,k)}$,
where $n_L(J,P,k)$ is the number of $L$ projectors contained in $P_{kk}(J^\cP)$, as seen in Table \ref{t2}.
Thus, the criterion for the absence of ghosts in the sector $(J^\cP)$ is that
\be
\underset{q^2=-m_n^2}{\mbox{Res}}\,
\sum_{k=1}^s \left[b^{-1}_{kk}(J^\cP) (-1)^{n_L(J,P,k)+1} \right] >0\ , \quad \forall{J,\cP, n}\ ,
\label{newgc}
\ee

\begin{table}[ht]
\begin{center}
\begin{tabular}{|c|c|c|c|c|}
\hline
   & $ts$ & $hs$ & $ha$ & $ta$ \\
\hline
$TTT$ & $3^-$, $1^-_1$ & $2^-_1$, $1^-_2$  & $2^-_2$, $1^-_3$ & $0^-$ \\
\hline
$TTL+TLT+LTT$  & $2^+_1$, $0^+_1$ & - & -  & $1^+_3$  \\
\hline
$\frac32 LTT$   & - & $2^+_2$, $0^+_2$  & $1^+_2$, & - \\
\hline
$TTL+TLT- \frac12 LTT$ & - & $1^+_1$ & $2^+_3$, $0^+_3$  & -  \\
\hline
$TLL+LTL+LLT$   & $1^-_4$ & $1^-_5$  & $1^-_6$   &  - \\
\hline
$LLL$    & $0^+_4$ & -  & -  &  - \\
\hline
\end{tabular}
\end{center}
\caption{$SO(3)$ spin content of a
three-index tensor. $ts/ta$=totally (anti)symmetric; $hs/ha$=hook (anti)symmetric.
The degrees of freedom relevant to Section 3 are contained in the first column, those relevant to Section 4 are contained in the first two columns.}
\label{t2}
\end{table}
\smallskip

A general rank-3 tensor carries the irreps listed
in Table \ref{t2}, that we reproduce from 
\cite{Percacci:2020ddy}.
The columns refer to the algebraic decomposition of
the tensor in its totally symmetric, hook symmetric, hook antisymmetric and totally antisymmetric parts.
The rows refer to the transverse/longitudinal character of the indices.
For a general rank three tensor, the spin projectors and intertwiners have been given in \cite{Percacci:2020ddy,Mendonca:2019gco}.
In this paper we only deal with symmetric tensors.

\section{ Model based on a totally symmetric tensor field}

\subsection{The quadratic action}

Let $Q_{\mu\nu\rho}=Q_{(\mu\nu\rho)}$ be a totally symmetric
third rank tensor of canonical dimension one, propagating in Minkowski space.
Since we will not discuss interactions, it is enough to
consider the most general Lagrangian containing terms quadratic in $Q_{\lambda\mu\nu}$:
\begin{align}
S = 
-\tfrac12 \int d^4 x\, \Big[ &
m_1Q_{\m\n\r}Q^{\m\n\r} +m_3Q_\mu Q^\mu
\nn\w2
& 
b_1 \big( \partial_\m Q_{\n\r\s}\big)^2 +b_3 (\partial_\m Q_\n )(\partial^\m Q^\n)+ b_6  \big(\divQ^{\m\n}\big)^2 + b_{10}\divQ^{\m\n} \partial_\m Q_\n
+ b_{14} (\partial_\m Q^\m)^2 
\Big]\ ,
\label{ga}
\end{align}
where 
\be
\divQ_{\m\n} := \partial^\r Q_{\r\m\n}\ , \qquad Q_\m := Q_{\m\n}{}^\n\ .  \ee
The first line contains the terms of dimension two,
and the second contain the terms of dimension four.
The reason for the non-consecutive numbering of the coefficients
will become clear in the next section, where we consider a more general
class of theories.

The case 
\be
m_1=m_3=0\ ,\quad
b_1=1\ ,\quad
b_3=-3\ ,\quad
b_6=-3\ ,\quad
b_{10}=6\ ,\quad
b_{14}=-3/2
\ee
corresponds to the Fronsdal action $S_F$, that describes only a massless spin-3 state.
It is invariant under the higher spin gauge transformation
\be
\delta Q_{\lambda\mu\nu}=\partial_{(\lambda}\Lambda_{\mu\nu)}\ ,
\label{hsgt}
\ee
where $\Lambda_{\mu\nu}$ is a symmetric traceless tensor. The corresponding coefficient matrices are given in (\ref{nograv2}), with the mass set to zero.

Returning to the general theory based on  \eq{ga}, we see from Table \ref{t2} that the field carries the following irreps:
\be
3^-,\ 2^+_1,\ 1^-_{1,4},\ 0^+_{1,4}\ .
\ee
Using the spin projector formalism, the quadratic action, in momentum space, can be written in the form
\bea
S^{(2)}&=&
\frac12 \int \frac{d^4 q}{(2\pi)^4}\ Q(-q)\bigg[
a(3^-)\,P(3^-)
+a_{11}(2^+)\,P_{11}(2^+)
\nonumber\\
&&\qquad\qquad\qquad
+\sum_{i,j=1,4} a_{ij}(1^-)\,P_{ij}(1^-)
+\sum_{i,j=1,4} a_{ij}(0^+)\,P_{ij}(0^+)
\bigg] Q(q) \ ,
\nonumber
\eea
where the matrices $a_{ij}(J^\cP,q)$ pertaining to the action (\ref{ga}), are 
\bea
&& a(3^-)=-q^2-m_1
\label{a3m}
\nn\w2
&& a(2^+)=
\left(b_1+\frac13 b_6\right)(-q^2)-m_1
\label{a2p}
\nn\w2
&& a(1^-) = 
\begin{blockarray}{cc}
\begin{block}{(cc)}
\left(b_1+\frac53 b_3\right)(-q^2) -m_1-\frac53 m_3  
&  \frac{\sqrt5}{6}\left(2b_3+b_{10}\right)(-q^2)-\frac{\sqrt5}{3}m_3
\\
&\\
\frac{\sqrt5}{6}\left(2b_3+b_{10}\right)(-q^2)-\frac{\sqrt5}{3}m_3 
&  \left(b_1+\frac13 b_3+\frac23b_6+\frac13b_{10}\right)(-q^2) 
-m_1-\frac13 m_3 \\
\end{block}
\end{blockarray}
\nn
\\
&& a(0^+) = 
\begin{blockarray}{cc}
\begin{block}{(cc)}
a(-q^2)-m_1-m_3 &  c(-q^2)-m_3   \\
&\\
c(-q^2)-m_3 &  b(-q^2)-m_1-m_3  \\
\end{block}
\end{blockarray}\ ,
\label{nograv}
\eea
where
\bea
a &:=& b_1+b_3+\tfrac13b_6+b_{14}\ ,
\nn\\
b &:=& b_1+b_3+b_6+b_{10}+b_{14}\ ,
\nn\\
c  &:=&  b_3+\tfrac12b_{10}+b_{14}\ .
\label{abc}
\eea
For generic values of the coefficients, they will be nondegenerate.
For special choices of the parameters in the Lagrangian,
there may be gauge symmetries lowering their rank.
Such symmetries are typically restricted to the quadratic action,
i.e. they do not correspond to symmetries of the full nonlinear action.
We want to avoid such accidental symmetries so that the ranks of the kinetic coefficient matrices are 1, 1, 2, 1 for spin $3^{-}, 2^{+}, 1^{-}, 0^{+}$, respectively. 

With this in mind the full source-to-source saturated propagator is 
\bea
\Pi &=& -\frac12\int \frac{d^4 q}{(2\pi)^4}\  \tau(-q) \Bigg[ a^{-1}(3^-)\,P(3^-)
+a^{-1}_{11}(2^+)\,P_{11}(2^+) 
\nn\w2
&& 
+\sum_{i,j=1,4} a_{ij}^{-1}(1^-)\,P_{ij}(1^-)
+\sum_{i,j=1,4} a_{ij}^{-1}(0^+)\,P_{ij}(0^+) \Bigg] \tau(q) \ .
\eea
\allowdisplaybreaks
For arbitrary values of the coefficients $b_i$, $m_i$, this propagator may, of course, harbor ghosts and/or tachyons.

\subsection{A no-go theorem}  

We are now going to prove the following statement:
There is no free Lagrangian for the totally symmetric field $Q_{\m\n\rho}$ that can describe the propagation of a massive spin-$3$ particle and any combination of lower spin particles in a manner that avoids ghosts and tachyons. 
This is almost the same as the statement of Singh and Hagen \cite{Singh:1974qz},
but here we do not insist that only spin-3 propagates.
Furthermore, we prove the statement using the spin projector formalism.

We assume that $3^-$ has a healthy propagation, implying that $b_1>0$ and $m_1>0$.
For the rest of this section, without loss of generality we will use the freedom of rescaling the field $Q$ to set
\be
b_1=1\ .
\label{cond2}
\ee
Then, we see from (\ref{a2p}) that $2^+$ necessarily has a mass with the wrong kinetic term. Thus we must inhibit its propagation, which implies
\be
b_6=-3\ .
\label{cond2}
\ee
Using this condition, we can rewrite
\bea
\det (1^-)& =& 
A_1q^4+B_1q^2
+m_1(m_1+2m_3)
\label{det1m1}
\\
&=&A_1\big(q^2-\mu^2_{1+} \big) \big(q^2-\mu^2_{1-}\big)\ ,
\label{det1m2}
\eea
where
\bea
A_1&=&-\frac{36+48b_3-12b_{10}+5b_{10}^2}{36}
\\
B_1&=&\frac{(b_{10}+6b_3)m_1-4m_3}{3}
\\
\mu_{1\pm}^2&=&
\frac{-B_1
\pm\sqrt\Delta_1}{2A_1}
\nonumber\\
\Delta_1&=&
B_1^2-4A_1 m_1(m_1-2m_3)
\nonumber\\
&=&
6m_1^2(6+b_{10}^2-2b_{10}+2b_3b_{10}+8 b_3+6b_3^2
)
\nonumber\\
&&\qquad
+\tfrac29 m_1m_3(36+24b_3-16b_{10}+5b_{10}^2)
+\tfrac{16}{9}m_3^2\ .
\eea
We now have to distinguish various cases. We begin by assuming $A_1\ne0$.
In this case we have two propagating $1^-$ states with squared masses $-\mu_{1\pm}^2$, that should be positive for consistency.
In particular, the product
\be
\mu_{1+}^2\mu_{1-}^2 = -\frac{36(m_1+2m_3)}{36+48b_3-12b_{10}+5b_{10}^2}\ .
\ee
should be positive.
These conditions can be satisfied. However, we also have to impose that there are no ghosts, which means  
\be
\lim_{q^2\to \mu_{1\pm}^2}(q^2-\mu_{1\pm}^2)\tr \,a(1^-)^{-1}<0\ .
\ee
as it follows from \eq{newgc}. These are rather complicated conditions, but using the Mathematica
command {\tt Reduce} we find that they cannot be satisfied simultaneously. Thus, one of the two states is a ghost, and this situation is ruled out.

Now consider the case where only one massive $1^-$ particle propagates.
To this end we set $A_1=0$ while keeping $\det a(1^-)\not=0$. 
\footnote{Note that in this case we cannot use expression (\ref{det1m2}) because $\mu_{1\pm}^2$ are singular.}
This gives
\be
b_3 = \frac{1}{48} \big(-36  + 12 b_{10} - 5 b_{10}^2 \big)\ .
\label{c1m1}
\ee
Then, using (\ref{det1m1}) we find that there is a single pole at
\be
q^2 = \frac{24m_1 (m_1+2m_3)} {(36- 20b_{10}+5 b_{10}^2)m_1+32m_3 }\equiv\mu_1^2\ .
\ee
One can arrange the coefficients such that the tachyon-free condition can be satisfied but the ghost-free condition is violated, since
\be
{\rm Res}\, \tr\,a(1^-)^{-1} |_{q^2=\mu_1^2} =24\frac{5(b_{10}-2)^2m_1^2+16(m_1+2m_3)^2}{(36-20b_{10}+ 5 b_{10}^2)m_1+32m_3)^2 }>0\ ,
\ee
so this case is pathological too. 

To remove this ghostly state as well, we now set to zero the coefficient of the $q^2$ term in \eq{det1m1}.
%
%
There are then two cases to be considered, depending on whether $m_1+2m_3$ is zero or not.
If $m_1+2m_3=0$, we have
\be
b_{10}=2\ ,\quad b_3=-\frac23\ .
\ee
However, in this case, $\det a(1^-)=0$  i.e. there is a 
gauge symmetry.
We fix the gauge by removing the second row and column of $a(1^-)$. The remaining nondegenerate submatrix is
\be
b(1^{-}) = \frac{q^2-\frac32 m^2}{9}\ ,
\ee
which shows that this state is tachyonic.   Thus the only remaining possibility for the $1^-$ states is that they do not propagate at all, and this will be true if $A_1=B_1=0$ and $m_1+2m_3\ne0$.
These conditions will be satisfied if (\ref{c1m1}) holds, and in addition
\be
m_3=-\frac{36-20b_{10}+5b_{10}^2}{32}m_1\ ,
\quad
\mathrm{and}
\quad
b_{10}\ne 2\ ,
\ee
the latter condition being necessary to avoid 
gauge symmetries, as we have already seen. 

Clearly, these conditions can be satisfied. We shall now see what consequences they have in the remaining $0^+$ sector.
The determinant is now
\bea
\det a(0^+)& =& 
A_0q^4+B_0q^2
-\frac{5}{16}(b_{10}^2-2)^2m_1^2
\label{det0p1}
\\
&=&A_0 \big(q^2-\mu_{0+}^2 \big) \big(q^2-\mu_{0-}^2\big)\ ,
\label{det0p2}
\eea
where
\bea
A_0&=&\frac{36-12b_{10}-b_{10}^2-48b_{14}}{24}
\\
B_0&=&\frac{-60 + 12 b_{10} + 5 b_{10}^2 + 96 b_{14}}{48}
m_1
\\
\mu_{0\pm}^2&=&
\frac{-B_0\pm\sqrt\Delta_0}{2A_0}
\nonumber\\
\Delta_0&=&B_0^2
+\frac54A_0(b_{10}^2-2)^2m_1^2
\eea
where 
\be
\mu_{0+}^2\mu_{0-}^2 = -\frac{15(b_{10}-2)^2}{-72 + 24 b_{10}+2b_{10}^2+96 b_{14}}m_1^2\ .
\ee
Again, to avoid tachyons we must have $\mu_{0+}^2<0$ and $\mu_{0-}^2<0$, and consequently $\mu_{0+}^2\mu_{0-}^2>0$. While  these conditions can be satisfied, requiring that there are no ghosts puts additional conditions which cannot be satisfied simultaneously. 

Next, we remove one of the spin $0^+$ states by setting $A_0=0$, leading to
\be
b_{14}=\frac{1}{48}
(36-12b_{10}-b_{10}^2)\ .
\ee
With this condition, the determinant becomes
\be
\det a(0^+)=\frac{1}{16}(b_{10}-2)^2 m_1(q^2-5m_1)\ .
\ee
Since this is a field of type TLT/LLL, the $m^2$ term has the wrong sign and is a tachyon.
We have thus exhausted all possible ways of having another particle propagate in addition to $3^-$, and all of them lead either to a tachyon or a ghost.

Thus, the only possibility that remains is to have $3^-$ propagate alone, 
which is well known to be impossible from \cite{Singh:1974qz}.
Within the spin projector formalism, this can be seen as follows.
Inhibiting the propagation of $2^+$ gives the condition (\ref{cond2}).
On the other hand, the elimination of all the $q^2$ dependent terms in $\det a(1^-)$ and $\det a(0^+)$ requires that
\be
b_{10}=2\ , \quad 
b_{14}= \frac16\ , \quad 
b_3= -\frac23\ , \quad  
m_3=-\frac12 m_1\ .
\label{s3}
\ee
However, for $m_3=-\frac12 m_1$ we get $\det a(1^-)=0$ and $\det a(0^+)=0$, which means that there are 
gauge symmetries. Using \eq{s3}, and removing the first row and column, which amounts to imposing gauge conditions to fix the 
gauge symmetry, we find
\bea
&& b(1^-) = \frac19 \left(q^2 -\frac32 m_1 \right)\ ,
\nn\w2
&& b^{-1}(0^+) = \frac12 \big( q^2 -m_1 \big)\ ,
\eea
and therefore both of them are tachyonic. This shows that it is not possible to propagate a unitary and causal spin $3^-$ state by the action \eq{ga},
neither alone nor together with some lower spin state.

\section{Inclusion of the hook symmetric tensor field}

\subsection{The quadratic action}

In view of the negative result of the preceding Section, we will now relax
the condition that the tensor field be totally symmetric
and only demand that it is symmetric in the second pair of indices:
\be
Q_{\lambda\mu\nu}=Q_{\lambda\nu\mu}\ .
\label{id}
\ee
In the rest of this section $Q_{\mu\nu\rho}$ is always understood to have only this symmetry. This may not seem to be a particularly natural generalization from the point of view of the theory of Lorentz covariant field equations,
but it is very natural in the context of MAG, as we have already discussed.
The most general Lagrangian for this field containing terms
of dimension two and four is 
\bea
\cL &=& 
-\frac12\Bigg[
m_1 \big(Q_{\mu\nu\rho}\big)^2
+m_2 Q^{\mu\nu\rho} Q _{\nu\mu\rho}
+m_3 (Q_\mu)^2
+m_4 (\tQ_\mu)^2
+m_5 Q_\mu \tQ_\mu 
\nn\w2
&&  
+b_1\big(\partial_\mu Q_{\nu\rho\sigma}\big)^2 
+b_2\big(\partial_\mu Q_{\nu\rho\sigma}\big)\big( \partial^\mu Q^{\rho\nu\sigma}\big)
\nn\w2
&& 
+b_3 \big(\partial_\mu \tQ_\nu \big)^2 
+b_4 \big(\partial_\mu Q_\nu \big)^2 
+b_5 \big(\partial_\mu \tQ_\nu \big)\big(\partial^\mu Q^\nu \big) 
\nn\w2
&& 
+b_6 \big(\divQ_{\mu\nu}\big)^2
+b_7\big(\divQt_{\mu\nu}\big)^2 
+b_8 \big(\divQt^{\mu\nu}\big)\divQt_{\nu\mu}
+b_9 \divQ^{\mu\nu}\divQt_{\mu\nu}
\nn\w3
&&  
+b_{10}\divQt^{\mu\nu} \partial_\mu \tQ_\nu 
+b_{11} \divQt^{\mu\nu} \partial_\mu Q_\nu 
+b_{12} \divQt^{\mu\nu} \partial_\nu \tQ_\mu 
+b_{13} \divQt^{\mu\nu} \partial_\nu Q_\mu 
\nn\w3
&& 
+b_{14} (\trdivQ)^2
+ b_{15} (\trdivQt)^2
+b_{16} \trdivQ \trdivQt 
\Bigg]\ ,
\label{tshs}
\eea
where we defined
\begin{align}
\divQ_{\mu\nu} &:=\partial^\lambda Q_{\lambda\mu\nu}\ ,
&&\divQt_{\mu\nu} :=\partial^\lambda Q_{\mu\nu\lambda}\ ,
\nn\w2
Q_\mu& :=g^{\lambda\tau} Q_{\mu\lambda\tau}\ ,
&&\quad\ \ \ \tQ_\mu :=g^{\lambda\tau} Q_{\lambda\tau\mu}\ ,
\nn\w2
\trdivQ& := g^{\mu\nu} \divQ_{\mu\nu}=\partial^\mu Q_\mu\ ,
&&\trdivQt := g^{\mu\nu} \divQt_{\mu\nu}=\partial^\mu\tQ_\mu\ .
\label{notation}
\end{align}
Thus we have $21$ parameters in total.
In the case when $Q_{\lambda\mu\nu}$ is totally symmetric,
many terms become identical and the number of parameters
reduces to $7$ (five $b$-type coefficients and two masses).
Without loss of generality, the Lagrangian can then be written as in (\ref{ga}).
The coefficient matrices $a_{ij}$ associated with the spin projection operators for the Lagrangian parametrized as displayed are given in Appendix
\ref{app:coefmat}.

From Table \ref{t2} we see that the $J^\cP$ sectors involved are
\be
3^{-},\ 2^+_{1,2},\ 2^{-},\ 1^{+},\ 
1^{-}_{1,2,4,5},\ 0^{+}_{1,2,4}\ . 
\label{euridice}
\ee
Before analyzing the spectrum of the states described by the Lagrangian \eq{tshs}, it will be useful to introduce an extension of the Fronsdal Lagrangian to the case of tensors that are only symmetric in two indices, that we shall call $\cL_{\rm EF}$.
It is characterized by the coefficients $b_i={\bar b}_i$ where

\def\bb{{\bar b}}

\begin{align}
\begin{tabular}{cccccccccccccccc}
$\bb_1$ & $\bb_2$ & $\bb_3$ & $\bb_4$ &  $\bb_5$ & $\bb_6$ & $\bb_7$ & $\bb_8$ &  $\bb_9$ & $\bb_{10}$ & $\bb_{11}$ &  $\bb_{12}$ & $\bb_{13}$ & $\bb_{14}$ & $\bb_{15}$ & $\bb_{16}$ \\
$\frac13$ & $\frac23$ & $-\frac43$ & $-\frac13$ & $-\frac43$ & $-\frac13$ & $-\frac23$ & $-\frac23$ & $-\frac43$ & $\frac83$ & $\frac43$ & $\frac43$ & $\frac23$ & $-\frac16$ & $-\frac23$ &
\begin{minipage}[c][7mm][t]{0.1mm}%
\end{minipage}
$-\frac23$\\
\end{tabular}
\label{bbar}
\end{align}
One arrives at these coefficients using the relation
\be
S_{\m\n\rho} := Q_{(\mu\nu\rho)} = \frac13\left(Q_{\m\n\rho} + 2Q_{(\nu\rho)\mu} \right)\ ,
\label{redef}
\ee
with $Q_{\m\n\rho}$ as defined in \eq{id}, in the Fronsdal Lagrangian. Thus we have
\be
{\cal L}_{\rm EF}(Q) = {\cal L}\big|_{m_i=0, b_i={\bar b}_i} = {\cal L}_F (S)\ ,
\label{diana}
\ee
with ${\cal L}$ from \eq{tshs}. This Lagrangian has the property that it describes only a massless spin $3^-$ particle and it reduces to the Fronsdal Lagrangian when the tensor $Q$ is totally symmetric.
In addition to the higher spin symmetry
\be
\delta_\Lambda Q_{\mu\nu\rho}=\partial_{(\rho}\Lambda_{\mu\nu)}
\label{hsgtEF}
\ee
it also has a shift invariance
\be
\delta_\xi Q_{\rho\mu\nu}=\xi_{\rho\mu\nu}
\label{hss}
\ee
where the parameter is hook symmetric: $\xi_{(\rho\mu\nu)}=0$.

The coefficient matrices of the extended Fronsdal Lagrangian are all zero, except for
\bea
a(3^{-}) &=& -q^2\ ,
\eea
\be
b(1^-)=
\begin{pmatrix}
4q^2 & 0 & 0 & 0\\
0 & 0 & 0 & 0\\
0 & 0 & 0 & 0\\
0 & 0 & 0 & 0
\end{pmatrix}
\label{EF_1m}
\ee

\be
b(0^+)=
\frac12
\begin{pmatrix}
9  & 0 & 3 \\
0 & 0 & 0 \\
3 & 0 & 1
\end{pmatrix}
q^2
\label{EF_0p}
\ee
We see again that all the hook symmetric degrees of freedom 
($2^+_2$, $2^-_1$, $1^-_{2,5}$, $0^+_2$) are pure gauge, and eliminating the corresponding rows and columns we indeed recover (\ref{nograv2}), with zero masses.
We will need these matrices later for comparison.

\subsection{Massive spin $3^{-}$ alone}

We will now seek appropriate sets of $b$-coefficients and masses such that only a massive spin 3 state propagates,
and such that there are no 
gauge symmetries.
To this end, we consider the general coefficient matrices, given in Appendix \ref{app:coefmat}.
The square mass of the spin-3 state is
$m^2=m_1+m_2$. We take it as reference mass and define the dimensionless mass ratios 
\be
\alsix=\frac{2m_1-m_2}{m^2}\ ,\quad
\alpha_i = \frac{m_i}{m^2} \qquad\mathrm{for}\qquad
i=3,4,5\ ,
\ee
so that the mass square of the spin $2^-$ and $1^+$ states are $\alpha m^2$.
Thus
\be
m_1=\frac{1+\alpha}{3}m^2\ ,\qquad
m_2=\frac{2-\alpha}{3}m^2\ .
\ee
The coefficient of $q^2$ in $a(3^-)$ is $-b_1-b_2$ and without loss of generality we can rescale the field so that $b_1+b_2=1$.
Then we see that in order to remove the $2^-$ and $1^+$ states we must demand
\be
b_1=\frac13\ ,\qquad
b_2=\frac23\ ,\qquad
b_7=b_8\ ,\qquad
\alsix\not=0\ ,
\ee
where the last condition ensures that there are no gauge symmetries.

In analyzing the spectrum of the model based on the Lagrangian \eq{tshs},  
the spin $1^{-}$ sector is the most involved one. To simplify matters, we first set $\alpha_4=\alpha_5=0$. Then we find a solution, that we call Model 1, with the parameters listed in the first column in Table 1. It contains only a massive $3^{-}$ state with non-tachyonic mass-squared $m^2$. 
\begin{table}[t!]
\small
\begin{center}
\begin{tabular}{|c|c|c|c|c|c|}
\hline
& Model 1 & Model 2 & Model 3  &  Model 4\\
\hline
$b_1$ & $1/3$ & $1/3$  & $1/3$ &   $1/3$ \\
\hline
$b_2$ & $2/3$ & $2/3$  & $2/3$ &   $2/3$ \\
\hline
$b_3$ & $-4/3$ & $-4/3$  & $-4/3$ &   $-4/3$ \\
\hline
$b_4$ & $-1/3$ & $-1/3$  & $-1/3$ &  $-1/3$ \\
\hline
$b_5$ & $-4/3$ & $-4/3$  & $-4/3$ &   $-4/3$ \\
\hline
$b_6$ & $-1/3$ & $-1/3$  & $-1/3$ &  $-1/3$ \\
\hline
$b_7$ & $-2/3$ & $-2/3$  & $b_7$ &   $-2/3$ \\
\hline
$b_8$ & $-2/3$ & $-2/3$  & $b_7$  &  $-2/3$ \\
\hline
$b_9$ & $-4/3$ & $-4/3$  & $-4/3$  &  $-4/3$ \\
\hline
$b_{10}$ & $8/3$ & $8/3$  & $8/3$ &  $8/3$ \\
\hline
$b_{11}$ & $4/3$ & $4/3$  & $4/3$ &  $4/3$ \\
\hline
$b_{12}$ & $4/3$ & $4/3$  & $4/3$ &   $4/3$ \\
\hline
$b_{13}$ & $2/3$ & $2/3$  & $2/3$ &  $2/3$ \\
\hline
$b_{14}$ & $11/9$ & $-11/72$  & $1/3$ &   $\left[3(2+3b_{16})^2-1\right]/6$ \\
\hline
$b_{15}$ & $-23/72$ & $2\big[3(2+3b_{16})^2-1\big]/3$ & $(1-4b_7)/8$ &  $-11/18$   \\
\hline
$b_{16}$ & $13/18$  & $b_{16}$ & $5/6$ &  $b_{16}$ \\
\hline
$\alpha_3$ & $5$ & $-2/9$  & $5/2$ &  $\left[(11+18b_{16})^2+2\right]/18$ \\
\hline
$\alpha_4$ & $0$ & $\big[ 2(11+18b_{16})^2-11\big]/9$ & $11/12$ &  $-8/9$   \\
\hline 
$\alpha_5$ & $0$ & $-14/9$  & $7$ &  $-20/9$ \\
\hline
$\alsix$ & $-15/2$ & $-1$  & $-1$ & $-4$ \\
\hline
\end{tabular}
\end{center}
\caption{Models which propagate solely a massive spin $3^{-}$ state. } 
\label{tab:mod}
\end{table}
We give here its coefficient matrices:
\bea
a(3^{-}) &=& -m^2 -q^2\ ,
\label{s3}\w2
a(2^{-}) &=& \frac{15}{4}m^2\ ,
\label{s2m}\w2
a(1^{+}) &=& \frac{15}{4}m^2\ ,
\eea
\be
b(1^-)=\frac13
\begin{pmatrix}
-28 & 10\sqrt5 & -5\sqrt5 & 5\sqrt 10\\
10\sqrt5 & -35/4 & 10 & -10\sqrt2\\
-5\sqrt5 & 10 & -8 & 5\sqrt2\\
5\sqrt 10 & -10\sqrt2 & 5\sqrt2 & 5/4
\end{pmatrix}
m^2
+
\begin{pmatrix}
4  & 0 & 0 & 0\\
0 & 0 & 0 & 0\\
0 & 0 & 0 & 0\\
0 & 0 & 0 & 0
\end{pmatrix}
q^2
\label{model1_1m}
\ee
\be
b(0^+)=
\begin{pmatrix}
-6 & -5\sqrt2 & -5 \\
-5\sqrt2 & -25/4 & -5\sqrt2 \\
-5 &  -5\sqrt2 & -6 
\end{pmatrix}
m^2
+
\frac{1}{8\sqrt2}
\begin{pmatrix}
11\sqrt2  & -25 & -13\sqrt2 \\
-25 & -25/2 & -25 \\
-13\sqrt2 & -25 & -21\sqrt2 
\end{pmatrix}
q^2
\label{model1_0p}
\ee
We observe that in the massless case the coefficient matrix of the spin $0^+$ sector does not have the extended Fronsdal form (\ref{EF_0p}). As a result, 
putting aside the mass terms, the remaining part of the Lagrangian does not have the extended Fronsdal form. Instead, we find
\bea
{\cal L}_{\rm Model\,1} &=& {\cal L}_{\rm EF}  
- \frac{1}{144} \left(100 (\trdivQ)^2 +25(\trdivQt)^2 +100\, \trdivQ\, \trdivQt\right) 
\nn\w2
&&
\qquad\qquad
-\frac12 m^2 \left[ -\frac{13}{6}\left(Q_{\mu\nu\rho}\right)^2
+\frac{19}{6} Q^{\m\n\rho} Q_{\n\rho\mu}+ 5\left(Q_\mu\right)^2 \right]\ ,
\eea
Without imposing that $\alpha_4$ and $\alpha_5$ vanish, we have found several more solutions, three of which (Models 2, 3, 4) are listed in Table 1.
Note that these are actually one-parameter families, with either $b_7$ or $b_{16}$ left unspecified.
All these models have no 
gauge symmetries,
as witnessed by the fact that the determinants of their coefficient matrices
$a(J^\cP)$ are nondegenerate. 
Note that, except for $b_7$ and $b_8$ in Model 3, the first 13 $b$-coefficients are equal to those of the extended Fronsdal Lagrangian.
The remaining three coefficients involve only terms formed with the scalar combinations $\rm trdivQ$ and $\rm trdiv\tilde Q$, so that their Lagrangians can be written as the extended Fronsdal one, plus mass terms, plus terms involving only these scalars:
\bea
{\cal L}_{\rm Model\,2} &=& {\cal L}_{\rm EF} 
-\frac{1}{144} (\trdivQ)^2
-9\left(b_{16}
+\frac23\right)^2(\trdivQt)^2
\nn\w2
&&
+ m^2\left[-\frac12Q^{\mu\nu\rho} Q _{\nu\rho\mu}+\frac19 (Q_\mu)^2-\frac16\left(77+264b_{16}+216b_{16}^2\right) (\tQ_\mu)^2  +\frac79 Q_\mu \tQ_\mu \right]\ ,
\nn\w2
{\cal L}_{\rm Model\,3} &=& {\cal L}_{\rm EF} 
-\left(b_7+\frac23\right) \tQ^{\mu\nu} \tQ_{(\mu\nu)} 
\nn\w2
&&
\qquad-\frac{1}{4} (\trdivQ)^2
-\frac{1}{48}\Big(19-12b_7\Big)(\trdivQt)^2
-\frac34 \, \trdivQ\, \trdivQt
\nn\w2
&&
\qquad-\frac12 m^2\Big[Q^{\mu\nu\rho} Q _{\nu\rho\mu} +\frac52 (Q_\mu)^2+ \frac{11}{2} (\tQ_\mu)^2  + 7 Q_\mu \tQ_\mu \Big]\ ,
\nn\w2
{\cal L}_{\rm Model\,4} &=& {\cal L}_{\rm EF} 
-\frac94 \left(b_{16}+\frac23\right)^2 (\trdivQ)^2
-\frac{1}{36}(\trdivQt)^2
-\frac12\left(b_{16}+\frac23\right) \, \trdivQ\, \trdivQt
\nn\w2
&&
\qquad +m^2\Big[\frac12Q^{\mu\nu\rho}Q_{\mu\nu\rho}
-Q^{\m\n\r} Q _{\nu\rho\mu} -\frac{1}{36}\left[\left(11+18b_{16}\right)^2+2\right](Q_\mu)^2 
 \nn\\
 && 
 \qquad\qquad+\frac49 (\tQ_\mu)^2+\frac{10}{9} Q_\mu \tQ_\mu \Big]\ .
\eea
Choosing the free parameters $b_{16}=-2/3$ and $b_7=-2/3$, i.e. the values they have in the extended Fronsdal Lagrangian, yields further simplifications giving 
\bea
{\cal L}_{\rm Model\,2}\Big|_{b_{16} =-2/3} &=& {\cal L}_{\rm EF} 
-\frac{1}{144} (\trdivQ)^2+ m^2\left\{-\frac12Q^{\mu\nu\rho} Q _{\nu\rho\mu}+\frac19(Q_\mu)^2 
+\frac12 (\tQ_\mu)^2
+\frac79 Q_\mu \tQ_\mu \right\}\ ,
\nn\w2
{\cal L}_{\rm Model\,3}\Big|_{b_7 =-2/3} &=& {\cal L}_{\rm EF} 
-\frac14 (\trdivQ)^2-\frac{9}{16} (\trdivQt)^2-\frac34  \, \trdivQ\, \trdivQt
\nn\w2
&& -\frac12m^2\Big[Q^{\mu\nu\rho} Q _{\nu\rho\mu} +\frac52 (Q_\mu)^2+ \frac{11}{2} (\tQ_\mu)^2  + 7 Q_\mu \tQ_\mu \Big]\ ,
\nn\w2
{\cal L}_{\rm Model\,4}\Big|_{b_{16} =-2/3} &=& {\cal L}_{\rm EF} 
-\frac{1}{36} (\trdivQt)^2
+ m^2\Big[\frac12Q^{\mu\nu\rho}Q^{\mu\nu\rho}-Q^{\m\n\rho} Q_{\nu\rho\mu}
\nn\w2
&&
-\frac{1}{12}(Q_\mu)^2+\frac49 (\tQ_\mu)^2+\frac{10}{9} Q_\mu \tQ_\mu \Big]\ .
\eea

It is noteworthy that the several local Lagrangians we have found all describe a spin $3^-$ state only. As we shall show in Section 4.6, these are not related to each other by any local field redefinition.  The kinetic matrices dressed with the spin projection operators are different for each of these models, and one can attribute the distinction between them to different mixings between the spin projections of the hook-symmetric field which essentially plays the role of auxiliary fields. Whether these Lagrangians can be written in a way that involves different set of independent auxiliary fields in a local manner remains to be investigated.

\subsection{Massless limits}

Even though our interest lies in massive theories, it is interesting to note that these models seem to become pathological in the massless limit.
As already remarked, in these models the coefficient matrices are nondegenerate, and there are no gauge symmetries.
This means that all the non-propagating degrees of freedom have Lagrangians that are pure mass terms. One would expect that when the masses are sent to zero, all these modes become pure gauge degrees of freedom. In other words, one would expect one gauge symmetry per unphysical degree of freedom.

However, the situation is a bit more complicated. 
For example, in model 1, the matrix element $a(1^-)_{11}$
is $4q^2$ plus a mass term, while all the others are purely mass terms, see Equation (\ref{model1_1m}). The reason why $q^2$ does not appear in the determinant (and so there is no propagation of $1^-$) is that the complementary minor has zero determinant. However, when the mass goes to zero,
\be
a(1^-)=
\begin{pmatrix}
4q^2&0&0&0\\
0&0&0&0\\
0&0&0&0\\
0&0&0&0\\
\end{pmatrix}
\ee
and so we apparently remain with a massless ghost.
Actually, this is non-propagating as it is needed to cancel some of the unphysical polarizations of the spin-3 state, as is seen from (\ref{EF_1m}).
The remaining unphysical polarizations should be removed by the spin $0$ terms.
However, this process does not go trough, because in the massless limit, the spin 0 coefficient matrix
\be
b(0^+)=
\frac{1}{8\sqrt2}
\begin{pmatrix}
11\sqrt2  & -25 & -13\sqrt2 \\
-25 & -25/2 & -25 \\
-13\sqrt2 & -25 & -21\sqrt2 
\end{pmatrix}
q^2
\label{model1_0plim}
\ee
does not have the required form (\ref{EF_0p}).
As a result, the massless limit of Model 1 contains a spin zero ghost, as can be seen explicitly by noting that the matrix in (\ref{model1_0plim}) has rank two, with one positive and one negative eigenvalue.

Analogous considerations hold also for all the other models.

\subsection{Contact terms}

Since the Lagrangians of our massive models do not have the form $\cL_{\rm EF}$+mass terms,
also the saturated propagator has a nonstandard form.
The standard saturated propagator for a $3^-$ state, following from Singh--Hagen theory, is given by (\ref{ms3}). In the extended theory where the $3^-$ state is carried by a tensor symmetric in two indices, this becomes
\bea
\Pi_{3m} &=& \frac12\int dq\, \tau_{\rm TS} \cdot \frac{P(3^-,m^2)}{q^2+m^2} \cdot \tau_{\rm TS}
\nn\\
&=& \frac12\int dq\, \frac{1}{q^2+m^2}\,\tau^{(\mu\nu\rho)} \left( P_{\mu\lambda}P_{\nu\tau}P_{\rho\sigma} -\frac35 P_{\mu\nu} P_{\lambda\tau} P_{\rho\sigma} \right) \tau^{(\lambda\tau\sigma)}\ ,
\label{ms3e}
\eea
where $\tau_{\rm TS}$ is the totally symmetric part of the source tensor $\tau$.
The explicit calculation of the saturated propagator (\ref{eve}) using the spin projectors (\ref{model1_1m},\ref{model1_0p}) yields, for Model 1,
\be
\Pi_{\rm Model 1} = \Pi_{3m}
+\frac{1}{10m^3}
\partial^\lambda\tau_{\lambda\mu}{}^\mu\left[\partial^\lambda\partial^\mu\partial^\nu\tau_{\lambda\mu\nu}+\frac18(q^2+7m^2)\partial^\sigma\tau_{\sigma\rho}{}^\rho
\right]
\ee
which becomes, on shell
\be
\Pi_{\rm Model 1} = \Pi_{3m}
+\frac{1}{10m^3}
\partial^\lambda\tau_{\lambda\mu}{}^\mu\left[\partial^\lambda\partial^\mu\partial^\nu\tau_{\lambda\mu\nu}+\frac34m^2\partial^\sigma\tau_{\sigma\rho}{}^\rho
\right]\ .
\ee
The additional terms represent contact interactions between the sources.
Similar contact terms are present also in the other models.

\subsection{A model with massive spin $3^{-}$ and $0^{+}$}

As an example in which some lower spin state propagates,
in addition to the $3^-$, we looked for models with a massive $0^+$.
Assuming, as before, that $\alpha_4=\alpha_5=0$ to simplify matters, 
one particular such model is given by choosing $b_i=\bar b_i$ for $i\not=14$,
with $b_{14}$ remaining as a free parameter, and
\be
\alsix= -10\ ,
\quad
\alpha_3=15\ ,\quad \alpha_4=\alpha_5=0\ ,
\ee
or equivalently
\be
m_1=-3 m^2\ ,\quad
m_2=4 m^2\ ,\quad
m_3=15 m^2\ ,\quad
m_4=m_5=0\ .
\ee
%
%
This choice of parameters gives
\bea
a(3^-)&=&-m^2-q^2\ ,
\\
a(2^-)&=&5m^2\ ,
\\
a(1^+)&=&5m^2\ ,
\\
\det[a(2^+)]&=&-5m^4\ ,
\\
\det[a(1^-)&=&625m^8\ ,
\\
\det[a(0^+)]&=&m^4\left(125m^2+\frac{24-71m^2}{3}q^2
\right)\ ,
\eea
showing that the propagating states are spin $3^-$ with mass $m^2$ and $0^+$ with mass 
\be
\mu^2(0^{+}) = \frac{375}{24b_{14} -71} m^2\ .
\ee
The Lagrangian takes the form
\be
{\cal L} = {\cal L}_{\rm EF} 
-\frac12\left(b_{14}+\frac16\right) \trdivQ^2  
+6m^2\Big[3Q^{\m\n\r} Q_{\m\n\r} -4Q^{\m\n\r} Q_{\n\m\r} 
-15Q^\mu Q_\mu
\Big]\ .
\label{tshs3}
\ee
The saturated propagator is
\be
\Pi=\Pi_{3m}
+\Pi_{0p}
+\frac{1}{75m^4}
\left[
5m^2\tau^{\mu\nu\rho}\tau_{\mu\nu\rho}
-8(\tilde \tau_\mu{}^{\mu\lambda}-\tau^{\lambda\mu}{}_\mu)\partial^\mu\partial^\nu(\tau_{\lambda\mu\nu}-2\tau_{\mu\nu\lambda})
\right]
\ee
where $\Pi_{3m}$ was given in (\ref{ms3e}) and
\be
\Pi_{0p}=
\frac12\int d^4q
J(-q)\frac{1}{q^2+\mu^2(0^+)}J(q)
\ee
is the spin $0^+$ saturated propagator, with the sources identified on shell as
\be
J=\frac{1}{10\sqrt5 m^3}\left(
10\partial^\mu\partial^\nu\partial^\rho\tau_{\mu\nu\rho}+2m^2{\rm trdiv}Q+\frac{12b_{14}-973}{24b_{14}-71}{\rm trdiv}\tilde Q
\right)\ .
\ee
We see that also in this case there are contact terms.
The spin $0^+$ state will have good propagation as long as $b_{14}>\frac{71}{24}$.

\subsection{Algebraic field redefinitions}
\label{fredef}

It is natural to ask whether the solutions listed in Table 1 are related to each other, or simpler solutions with more parameters set to zero can be obtained by algebraic field redefinitions. Consider the redefinition
\be
Q_{abc}\mapsto Q_{abc}+ \beta_1Q_{(bc)a} + \beta_2 Q_{(b} g_{c)a}  +\beta_3 \tQ_{(b} g_{c)a} +\beta_4 Q_a g_{bc}+\beta_5\tQ_a g_{bc}\ ,
\label{qredef}
\ee
where $\beta$'s are constants, and the notations used are defined in \eq{notation}. Using the spin projector operators provided in \cite{Percacci:2020ddy}, this field redefinition can be expressed as
\be
Q_{abc} \to  \sum_{ i,j,J^\cP}\beta_{ij} P_{ij} (J^\cP)_{abc}{}^{def} Q_{def}\ ,
\label{rd}
\ee
where $\beta_{ij}$ is a constant matrix. A single condition on the parameters $\beta_i$ needs to be imposed such that $\det \beta\ne 0$, to ensure the invertibility of the field redefinition. For an explicit example, see Appendix B in the case of spin $2^+$. Performing the above field redefinition in \eq{tshs}, the resulting field equation is 
\be
\sum_{ i,j,J^\cP} (\beta^T a \beta)_{ij} P_{ij} Q = \sum_{ i,j,J^\cP}\beta_{ij} P_{ij} (J^\cP) J\ ,
\label{rd5}
\ee
where the contractions of the triple of indices as in \eq{rd} are suppressed to avoid clutter in notation. Solving for $P_{ij}Q$ and substituting back into \eq{tshs} gives the saturated propagator in the momentum space as
\be
\Pi = \int dq\, J \cdot a_{ij}^{-1} P_{ij}(J^\cP) \cdot J\ ,
\ee
where $a_{ij}$ are the coefficient matrices in the $J^\cP$ sector, as dictated by the values of $\beta_i$, which in turn are determined by the requirement that \eq{tshs} describes a sole massive spin $s^-$ state. Thus, the algebraic field redefinition do not change the spectrum, as expected. Using the field redefinition (\ref{rd5}), we have also checked that none of the solutions listed in Table \ref{tab:mod} can be related to each other for any choice of the parameters $\beta_i$. As to whether the redefinition \eq{rd5} can be used to set to zero a set of parameters $b_j, a_k$, while we do not have a conclusive result about the most general possible outcome, in practice we have found it to be highly constraining requirement, with no solution we can offer here.

\section{Discussion }
\label{discussion}

\noindent{\it Search strategies}\\
When one looks for Lagrangians that propagate only healthy degrees of freedom, there are two ways of getting rid of unwanted modes. The first is to choose the parameters in the Lagrangian in such a way that when one diagonalizes the kinetic coefficients $a(J^\cP)_{ij}$, the entry corresponding to the unwanted field does not contain $q^2$, in other words its quadratic Lagrangian is just a mass term.
The equation of motion coming from such a Lagrangian forces the unwanted field to be zero.
The second way is to choose the parameters in the Lagrangian so that not only the $q^2$ term is absent, but also the mass. The quadratic Lagrangian for such a degree of freedom is then zero, and we say that the degree of freedom is pure gauge.

Each of these two methods comes with some danger.
The danger of imposing gauge invariance at the quadratic level is that it may turn out to be an accidental invariance, which means that it may be impossible, to construct invariant interactions \cite{Bekaert:2010hw}. In our context, the prime example is the tensor symmetry of the Fronsdal Lagrangian, whose nonlinear completion has never been constructed.
On the other hand, suppressing the propagation of some degrees of freedom by mass Lagrangians, depends on choices of coefficients that may not be respected by the radiative corrections, once one tries to quantize the theory.
For this reason, the authors of \cite{Barker:2025xzd,Barker:2025rzd,Marzo:2021iok,Marzo:2024pyn,Barker:2025fgo} advocate the use of gauge symmetries to eliminate unwanted degrees of freedom.
In this paper we have chosen to avoid accidental symmetries, so the stability of the results under radiative correction is an open issue that remains to be investigated.

\smallskip

\noindent{\it Spin 3 interactions from the bottom up}\\
Starting from quadratic action we have considered in this paper, the interactions can be introduced by various means.  Considering the interactions of massive spin 3 to lower spin fields, the minimal coupling of gravity to the Singh-Hagen description of the massive spin 3 has been considered by many authors, see for example \cite{Cucchieri:1994tx} and references therein. We shall come back to this point below. The need to improve their unitarity behaviour by adding higher derivative terms to the action has been pointed out in \cite{Cucchieri:1994tx}.  

Another approach to the construction of the interactions order by order in number of fields is the Noether procedure. The complexities in this approach have been discussed in \cite{Boulanger:2018dau}, where the advantages of the Batalin-Vilkovisky (BV) formalism was pointed out and applied to the construction of the interactions of massive spin 2 field. 
In this approach one works with a gauge invariant formulation of the massive theory in which the gauge symmetries are St\"uckelberg--type shift symmetries. Such a formulation exists also for the Singh-Hagen system for massive spin 3 \cite{Zinoviev83} by introducing the extra fields $(S_{\mu\nu}=S_{\nu\mu}, A_\mu, \varphi)$. \footnote{Note that the linearized level symmetries explored in \cite{Barker:2025rzd,Barker:2025fgo} do not involve extra fields.}
This model is recalled in Appendix B for the reader's convenience. The problem of interactions in this model was considered by Zinoviev \cite{Zinoviev:2008ck} who showed that such interactions exist consistently for Ricci flat backgrounds. 
The coupling of massive spin 3 to lower spins have also been studied in \cite{Bellazzini:2019bzh} focusing on the issue of high energy unitarity.

When interactions are turned on, it has been long known that the coupling of massive fields with spin $s>2$ to gravity and massive or massless fields with spin $s<2$ suffer breakdown in unitarity at energies ${\sqrt s}\sim \sqrt{m M_{\rm Pl}}$, which can be much below the Planck scale $M_{\rm Pl}$ for small $m$. It has been shown that this behaviour can be improved by adding non-minimal couplings which depend on curvatures and their derivatives~\cite{Ferrara:1992yc,Porrati:1993in,Cucchieri:1994tx}.

\smallskip

\noindent{\it Spin 3 interactions from MAG}\\
In the case of spin 3, MAG offers an alternative ``top down" approach.
MAG is typically presented as a theory with independent metric $g$ and connection $A$.
We call this the ``Cartan" form of MAG.
Since we are mostly interested in spin 3, we can neglect torsion, so the feature that distinguishes $A$ from the Levi-Civita connection $\mit{\Gamma}$ is the nonmetricity 
\be
Q_{\lambda\mu\nu}=-D_\lambda g_{\mu\nu}
\equiv -\partial_\lambda g_{\mu\nu}
+A_\lambda{}^\rho{}_\mu g_{\rho\nu}
+A_\lambda{}^\rho{}_\nu g_{\mu\rho}\ .
\label{nonmet}
\ee
The difference between $A$ and $\mit{\Gamma}$ is a tensor
\be
L_\mu{}^\rho{}_\nu
=A_\mu{}^\rho{}_\nu-\mit{\Gamma}_\mu{}^\rho{}_\nu
\ee
that is sometimes called the distortion or disformation.
It is related to the nonmetricity by
\be
L_{\lambda\mu\nu}= \frac{1}{2}\left(Q_{\lambda\mu\nu}+Q_{\nu\mu\lambda}-Q_{\mu\lambda\nu}\right)\ .
\label{nonmet}
\ee
The action $S(g,A)$ is an integral of monomials formed with
$L$, $F$ (the curvature of $A$) and their $A$-covariant derivatives
$DL$, $DF$ etc.
In this form, the theory looks very much like other gauge theories,
and one can argue that it is in a Higgs phase,
in the sense that the connection is massive,
with the metric acting like an order parameter.
 
There exists another form of the theory, that we call the ``Einstein" form.
It consists of taking $g$ and $L$ as independent variables,
or equivalently $g$ and $Q$.
The action $S(g,Q)$ is then an integral of monomials formed with
$Q$, $R$ (the Riemann tensor) and their Levi-Civita covariant derivatives
$\nabla Q$, $\nabla R$ etc.
In this form, the theory looks like ordinary metric gravity
coupled to a tensorial ``matter" field $Q$.
It is this form that is best suited for the discussion of spin 3,
since $Q$, now viewed as an independent variable,
can be immediately identified with the field that carries spin 3 in higher spin theories.

The question of forming interaction terms now looks different.
The interaction of spin 3 with gravity, limited to terms of dimension up to 4,
will involve the minimal covariantization of (\ref{tshs})
\bea
\cL &=& 
-\frac12\sqrt{|g|}\Bigg[
m_1 \big(Q_{\mu\nu\rho}\big)^2   
+m_2 Q^{\mu\nu\rho} Q _{\nu\mu\rho}
+m_3 (Q_\mu)^2
+m_4 (\tQ_\mu)^2
+m_5 Q_\mu \tQ_\mu 
\nn\w2
&&  
+b_1\big(\nabla_\mu Q_{\nu\rho\sigma}\big)^2 
+b_2\big(\nabla_\mu Q_{\nu\rho\sigma}\big)\big( \nabla^\mu Q^{\rho\nu\sigma}\big)
\nn\w2
&& 
+b_3 \big(\nabla_\mu \tQ_\nu \big)^2 
+b_4 \big(\nabla_\mu Q_\nu \big)^2 
+b_5 \big(\nabla_\mu \tQ_\nu \big)\big(\nabla^\mu Q^\nu \big) 
\nn\w2
&& 
+b_6 \big(\divQ_{\mu\nu}\big)^2
+b_7\big(\divQt_{\mu\nu}\big)^2 
+b_8 \big(\divQt^{\mu\nu}\big)\divQt_{\nu\mu}
+b_9 \divQ^{\mu\nu}\divQt_{\mu\nu}
\nn\w3
&&  
+b_{10}\divQt^{\mu\nu} \nabla_\mu \tQ_\nu 
+b_{11} \divQt^{\mu\nu} \nabla_\mu Q_\nu 
+b_{12} \divQt^{\mu\nu} \nabla_\nu \tQ_\mu 
+b_{13} \divQt^{\mu\nu} \nabla_\nu Q_\mu 
\nn\w3
&& 
+b_{14} Q^2
+ b_{15} \tQ^2
+b_{16} Q \tQ 
\Bigg]\ ,
\label{tshscov}
\eea
where now $\divQ_{\mu\nu}=\nabla^\rho Q_{\rho\mu\nu}$ etc. There are ambiguities in the covariantization process due to integration by parts and noncommutativity of the covariant derivatives. The consequences of these ambiguities in the context of tree level unitarity has been discussed in \cite{Cucchieri:1994tx}. 
To this Lagrangian one should add various other terms: the (dimension two) Hilbert term $m_P^2 R$, the (dimension four) terms of quadratic gravity $\alpha R^2+\beta R_{\mu\nu}R^{\mu\nu}
+\gamma R_{\mu\nu\rho\sigma}R^{\mu\nu\rho\sigma}$ and the mixing terms
\begin{equation}
    \begin{aligned}
H^{RQ}_{4} &= R^{\alpha\beta} \divQ_{\alpha\beta} ~, &
\quad
H^{RQ}_{5} &= R^{\alpha\beta} \divQt_{\alpha\beta} ~, &
\\
H^{RQ}_{6} &= R \, Q ~, &
\quad
H^{RQ}_{7} &= R \, \tQ ~. &
\label{HRQ}
\end{aligned}
\end{equation}
All these terms, when expanded around flat space,
\be
g_{\mu\nu}=\eta_{\mu\nu} \quad\mathrm{and}\quad Q_{\rho\mu\nu}=0\ ,
\ee
are quadratic in the metric fluctuation $h_{\mu\nu}$ and $Q_{\lambda\mu\nu}$, and thus contribute to the propagator. Note that the terms (\ref{HRQ}) could in principle give rise to mixing between $h_{\mu\nu}$ and the spin-2 parts of $Q_{\lambda\mu\nu}$.
However, such terms can be eliminated by the field redefinitions discussed in Section \ref{fredef}, so that without loss of generality one can assume that $h_{\mu\nu}$ and $Q_{\lambda\mu\nu}$
have independent propagators.

We digress briefly to comment on \cite{Baekler:2006vw},
who found spin 3 in a gauge theory of gravity with Lagrangian quadratic in the curvature $F$. When converted to the Einstein form, the Lagrangian contains both torsion and nonmetricity, but the quadratic terms depend only on nonmetricity. There is an accidental shift symmetry that removes torsion from the linearized theory.
We show in Appendix \ref{bbh} that their quadratic Lagrangian agrees with our extended Fronsdal Lagrangian $\cL_{\rm EF}$, up to a linear field redefinition. The reason for this is that the terms of the form $F^2$ do not form a complete basis of invariants, as discussed in \cite{Baldazzi:2021kaf}. In order to get $\cL_{\rm EF}$ one would have to complement the $F^2$ terms with some terms of the form $(DQ)^2$.

Finally, there are further dimension-four terms
of the schematic form $QQR$, $QQ\nabla Q$, $QQQQ$
that, once expanded around flat space, only contribute to the vertices.
In particular, these terms contain the cubic and quartic self-interactions
of the spin 3 degrees of freedom. There are altogether close to 200 terms that one can have in a Lagrangian of this type.
Thus, one could start by making a catalogue of all the possible independent interaction terms
and then seek subclasses
that would have all the desired good properties, at least in some effective field theory regime.
One attractive feature of this approach is that MAG, in its Cartan form, offers a geometrization of the spin 3 sector in the tower of higher spin fields that are expected to arise to achieve high energy unitarity.


Once the interactions are added to any of the models we have introduced, it will be interesting to study the resulting perturbation theory around Minkowski spacetime. Different models describe spin 3 state only but differ in the nature of the mixings between different projections of the hook symmetric field. In the construction of interactions this may lead to different interactions and as such it would be useful to explore the consequences in the computation of the amplitudes.

\section*{Acknowledgment}

R.P. wishes to thank W. Barker, C. Marzo and A. Santoni for useful correspondence. The work of E.S. is supported in part by NSF grant PHY-2112859 and PHY-2413006. 

\newpage

\begin{appendix}

\section{The propagators for massless and massive spin $0^\pm, 1^\pm, 2^\pm$ }
\label{standardpropagators}

In metric affine gravities considered here, the propagating modes with spin $0^\pm, 1^\pm, 2^\pm ,3^- $ are sourced by various projections of the sources $(\tau_{cab},\sigma_{ab})$ for the fields $(A_{cab}, h_{ab})$.  The purpose of this appendix is to review the description of these spins in terms of minimal rank tensors when available.
In finding the propagators of these fields it is convenient to introduce the notation
\be
P(J^\cP,\mu^2)= P(J^\cP)\Big|_{q^2=-\mu^2}\ ,\quad
P(J^\cP,\eta)=P(J^\cP)\Big\vert_{\partial\to 0}\ ,
\quad
P_{\mu\nu} = \left(\eta_{\mu\nu} + \frac{q_\mu q_\nu}{\mu^2} \right)\ .
\label{d1}
\ee
where $\mu$ is the mass of the state under consideration.

\subsection*{Spin $0^\pm$ }
%

\subsubsection*{Massive spin $0^\pm$}

In our conventions, a ghost and tachyon free massive spin $0^+$ field action takes the form
\bea
S(0^+) &=& \int d^4 x\,  \left[ \frac12 \phi (\Box-m^2)  \phi  +J\phi \right]\ ,
\w2
S(0^-) &=& \int d^4 x\,  \left[ -\frac12 \phi^{\mu\nu\rho}  (\Box-m^2)  \phi_{\mu\nu\rho}  +J^{\mu\nu\rho} \phi_{\mu\nu\rho} \right]\ ,
\label{sm}
\eea
where $\phi_{\m\n\rho}= \phi_{[\m\n\rho]}$ and $J_{\m\n\rho}= J_{[\m\n\rho]}$.  It follows that the source to source propagators in momentum space are given by
\bea
\Pi(0^+)  &=& \frac12 \int dq\,J(-q)\,\frac{1}{q^2+m^2}\, J(q)\ ,
\label{sm2}\w2
\Pi(0^-) &=& \frac12 \int dq\, J^{\mu\nu\rho}(-q)\, \frac{1}{q^2+m^2}\, P_{\mu\nu\rho}{}^{\lambda\rho\sigma} (0^-,m^2)\,J_{\lambda\tau\sigma}(q)
\nn\w2
&=& \frac12 \int dq\, J^{\mu\nu\rho}(-q)\, \frac{1}{q^2+m^2}\, P_{\mu\lambda} P_{\nu\tau} P_{\rho\sigma}\,J_{\lambda\tau\sigma}(q)
\eea
The massive spin $0^-$ can also be described in terms of a vector field by the following action
\begin{align}
S &=\int d^4x\, \Big[ -\frac12\left(\partial_\mu A^\mu\right)^2 - \frac12 m^2 A_\mu A^\mu + J^\mu A_\mu \Big]
\nn\w2
&=\int dq \Big\{ -\frac12 A \cdot \Big[ (q^2+m^2) P^0 + m^2 P^1\big]\cdot A+J\cdot A \big\} \ .
\end{align}
which gives the saturated propagator
\begin{align}
\Pi & = \int dq\left[\frac12 J(-q) \cdot \frac{P^0}{q^2+m^2} \cdot J(q) +\frac{1}{2m^2} J(-q)\cdot P^1 (-q) \cdot J(q)\right]
\nn\w2
&= \int dq \left[\frac12 J^\mu (-q) \frac{1}{q^2+m^2} \frac{q_\mu q_\nu}{m^2} J^\nu(q) +\frac{1}{2m^2} J_\mu J^\mu\right]\ .
\end{align}

\subsubsection*{Massless spin $0^\pm$}

For the massless spin $0^\pm$ fields the propagators are
\bea
\Pi(0^+)  &=& \frac12 \int dq\,J(-q)\,\frac{1}{q^2}\, J(q)\ ,
\nn\w2
\Pi(0^-) &=& \frac12\int dq\, J^{\mu\nu\rho}(-q)\, \frac{1}{q^2}\, P_{\mu\nu\rho}{}^{\lambda\rho\sigma} (0^-,\eta)\,J_{\lambda\tau\sigma}(q)
\nn\w2
&=& \frac12 \int dq\, J^{\mu\nu\rho}(-q)\, \frac{1}{q^2}\, J_{\mu\nu\rho}(q)
\eea
In the vector field formulation, taking $m^2=0$ in the action yields the saturated propagator
\be
\Pi= \frac12 \int dq\, J \frac{P^0}{q^2} \cdot J = \frac12 \int dq \,\frac{J\cdot J}{q^2} \ ,
\ee
with the source constraint $P^1\cdot J=$ that follows from the equation of motion.

\subsection*{Spin $1^-$}

\subsubsection*{Massive spin $1^-$ }

The ghost and tachyon free  massive spin $1^-$ field Lagrangian is that of Proca, namely 
\begin{align}
S & = \int d^4x \Big(-\frac14 F_{\mu\nu} F^{\mu\nu}
-\frac12 m^2 A_\mu A^\mu +J^\mu A_\mu\Big)\ .
\end{align}
The Fourier transformed action is
\begin{align}
S &=\int dq \Big\{-\frac12 A\cdot  \Big[(q^2+m^2)P^1 +m^2 P^0\Big]\cdot A +J \cdot A\Big\}\ ,
\end{align}
where $P^0, P^1$ are the standard spin $0, 1$ projectors
\be
P^1_{\mu\nu} = \eta_{\mu\nu} - \frac{q_\mu q_\nu}{q^2}\ ,\qquad P^0_{\mu\nu}=\frac{q_\mu q_\nu}{q^2}\ .
\ee
It follows that the saturated propagator is 
\bea
\Pi  &=& \frac12\int dq\ J_\mu(-q) \left[ \frac{1}{q^2+m^2} P(1^-,m^2)^{1\mu\nu} \right] J_\nu\ .\,
\nn\w2
&=& \frac12\int dq\ J^\mu(-q)\,\frac{P_{\mu\nu}}{q^2+m^2}\,J^\nu(q)\ ,
\label{ppp}
\eea
with the definition \eq{d1} understood. 

\subsubsection*{Massless spin $1^-$}

In this case, using the source constraint $q^\mu J_\nu$, the saturated propagator is
\bea
\Pi  &=& \frac12\int dq\ J_\mu(-q)  \frac{1}{q^2} P^{1\mu\nu}  J_\nu\ .\,
\nn\w2
&=& \frac12\int dq\ J_\mu(-q)\,\frac{1}{q^2}\eta^{\mu\nu}\,J_\nu(q)\ ,
\label{ppp}
\eea

\subsection*{Spin $1^+$}

\subsubsection*{Massive spin $1^+$ }

Massive spin $1^+$ state can be described by a second rang antisymmetric tensor $B_{\mu\nu}$ with the action
\begin{align}
S &= \int d^4x\, \Big( -\frac16 H_{\mu\nu\rho}H^{\mu\nu\rho} -\frac12 m^2 B_{\mu\nu}B^{\mu\nu}+ J^{\mu\nu} B_{\mu\nu}\Big)\ ,
\label{1pa}
\end{align}
where $H_{\mu\nu\rho} = 3\partial_{[\mu} B_{\nu\rho]}$. Using the spin projectors 
\be
P_a(1^+)_{\mu\nu}{}^{\rho\sigma} = P^1_{[\mu}{}^\rho P^1_{\nu]}{}^\sigma\ ,\qquad P_a(1^-)_{\mu\nu}{}^{\rho\sigma} = P^1_{[\mu}{}^\rho P^0_{\nu]}{}^\sigma-P^1_{[\mu}{}^\sigma P^0_{\nu]}{}^\rho\ ,
\ee
the Fourier transformed action
\begin{align}
S=\int d^4 q\Big\{ -\frac12 B\cdot \Big[(q^2+m^2)P_a(1^+)+m^2P_a(1^{-})\Big]\cdot B + J\cdot B\Big\}\ .
\label{1pa}
\end{align}
Solving for $B_{\mu\nu}$ in terms of the source and substituting back into the action gives the saturated propagator
\begin{align}
\Pi &=\frac12 \int \frac{d^4q}{(2\pi)^4} J(-q)\cdot \Big[\frac{1}{q^2+m^2} P_a(1^+)+\frac{1}{m^2} P_a(1^{-})\Big]\cdot J(q)
\nn\w2
&= \frac12 \int \frac{d^4q}{(2\pi)^4} J(-q) \frac{P(1^+,m^2)}{q^2+m^2} \cdot J(q)\ ,
\end{align}
where $P(1^+,m^2) := P_a(1^+,q^2)\Big|_{q^2=-m^2}$, describing 3 degrees of freedom. Upon dualization it can be seen that the massive spin $1^+$ state described by the action \eq{1pa} is dual to a massive spin $1^-$ field described a vector field described by the usual Proca action \cite{Sezgin:2012ka}.

\subsubsection*{Massless spin $1^+$ }

Setting $m=0$ in \eq{1pa} gives
\begin{align}
\Pi &=\frac12 \int \frac{d^4q}{(2\pi)^4} J(-q)\cdot \Big[\frac{1}{q^2+m^2} P_a(1^+)\Big]\cdot J(q)
\nn\w2
&= \frac12 \int \frac{d^4q}{(2\pi)^4} J^{\mu\nu}(-q)\frac{1}{q^2} J_{\mu\nu}(q)\ ,
\end{align}
where the constraint $P_a(1^{-})\cdot J=0$, which follows from the field equation, has been used. As is well known, the action for massless $B_{\mu\nu}$ on-shell describes a massless scalar as can easily be seen upon dualization.

\subsection*{Spin $2^+$} 

\subsubsection*{Massive spin $2^+$ }

Adding Fierz-Pauli mass term to the linearized Einstein Lagrangian gives
the action 
\begin{align}
S &= \int d^4x\,\Big( -\frac12 h^{\mu\nu} \Box h_{\mu\nu} + \frac12 h\Box h +(\partial_\mu h^{\mu\nu}) \partial_\nu h - (\partial_\mu h^{\mu\nu}) \partial^\rho h_{\nu\rho} + \frac12 m^2 (h_{\mu\nu} h^{\mu\nu} -h^2) + J^{\mu\nu} h_{\mu\nu}\Big)\ .
\label{Einstein}
\end{align}
The resulting Fourier transformed action is
\begin{align}
S &= \frac12 \int dq \Big\{ (q^2+m^2) h\cdot \left[ P(2^+)_{44}-2P(0^+)_{55} \right]\cdot h
\nn\w2
& +m^2 h\cdot \left[P(1^-)_{77} -{\sqrt 3} P(0^+)_{56} -{\sqrt 3} P(0^+)_{65}\right]  \cdot h +J\cdot h \Big\}\ ,
\label{m2a}
\end{align}
which yields the saturated propagator
\bea
\Pi &=& \int dq\, J (-q) \cdot \Big[ \frac{1}{q^2+m^2} P(2^+,q)_{44}+ \frac{1}{m^2} P(1^-,q)_{77} 
\nn\\
&& + \frac{2}{3m^4} (q^2+m^2) P(0^+,q)_{66} -\frac{1}{{\sqrt 3}m^2} P(0^+,q)_{56} -\frac{1}{{\sqrt 3}m^2}  P(0^+,q)_{65}  \Big] \cdot J(q)
\eea
Substituting the expressions for the spin projection operators gives 
\bea
\Pi (2^+) &=& \int dq\, J (-q)\cdot\, \frac{P(2^+,m^2)_{44}}{q^2+m^2}\cdot\, J(q)
\w2
&=& \int dq\, J^{\mu\nu} (-q)  \frac{1}{q^2+m^2} \left( P_{\mu\rho} P_{\nu\sigma} -\frac13 P_{\mu\nu} P_{\rho\sigma} \right)  J^{\rho\sigma}(q)\ .
\eea

\subsubsection*{Massless spin $2^+$ }

Turning to the case of massless spin 2 field, it follows from \eq{m2a} that the saturated propagator is then given by
\be
\Pi = \int dq\, J (-q)\cdot \frac{1}{q^2}\left( P(2^+,q)_{44} - \frac12 P(0^+,q)_{55}\right) \cdot J(q)\ .
\label{m2b}
\ee
It also follows from \eq{m2a} that the sources obey the constraints
\be
P(1^-)\cdot J=0\ ,\qquad P(0^+)_{66}\cdot J=0\ ,\qquad P(0^+)_{56}\cdot J=0\ .
\ee
These constraints are equivalent to $q^\mu J_{\mu\nu}(q)=0$. Using this constraint in \eq{m2b}, one readily finds that
\bea
\Pi(2^+)  &=& \int dq\, J (-q)\cdot \frac{1}{q^2}\, \left(P(2^+,\eta)_{44} -\frac12 P(0^+,\eta)_{55} \right)\cdot J(q) 
\nn\w2
&=& \int dq\, J^{\mu\nu} (-q) \frac{1}{q^2}\left( \eta_{\mu\rho}\eta_{\nu\sigma}-\frac12 \eta_{\mu\nu}\eta_{\rho\sigma}\right) J^{\rho\sigma}(q)\ ,
\label{m2c}
\eea
where we have used \eq{d1}.

\subsection*{Spin $2^-$} 

\subsubsection*{Massive spin $2^-$ }

A free action for a massive spin $2^-$ particle, in terms of a hook symmetric 3rd rank tensor field defined by
\be
\phi_{cab}= \phi_{c(ab)}\ ,\quad \phi_{(abc)}=0\ ,
\label{hsd}
\ee
is given by 
\begin{align}
  S &=\int d^4 x \Big[ -\frac12 (\partial_\mu \phi_{\nu\rho\sigma})^2
  +3(\partial_\mu \tilde\phi_\nu)^2
  -3 (\partial_\mu \tilde\phi^\mu)^2
-6(\partial^\mu\tilde\phi^\nu ) \partial^\rho\phi_{\mu\rho\nu}
\nn\w2
& +\frac12\left(\partial^\rho \phi_{\rho\mu\nu}\right)^2
+\left(\partial^\rho \phi_{\mu\rho\nu}\right)^2
-\frac12 m^2 (\phi_{\mu\nu\rho})^2
+3 m^2 (\tilde\phi_\mu)^2 \Big]\ ,
\end{align}
where $\phi_\mu := \phi_{\mu\nu}{}^\nu$, $\tilde\phi_\mu:=\phi^\rho{}_{\rho\mu}$.

The Fourier transform of the action is given by
\begin{align}
S &=\int dq \Big\{ -\frac12 \phi\cdot (q^2+m^2)\left[P_{11}(2^-)-P_{22}(1^-)\right]\cdot \phi +J\cdot \phi
\nn\w2
&-\frac12 m^2 \phi\cdot\left[P_{11}(1^+)+P_{22}(2^+) -2P_{22}(0^+) -\sqrt2 P_{25}(1^-)-\sqrt2 P_{52}(1^-)\right] \cdot \phi \Big\}\ ,
\label{a2m}
\end{align}
and the saturated propagator by
\be
\Pi= \frac12 \int \frac{d^4q}{(2\pi)^4} J(-q)\cdot \frac{1}{q^2+m^2} P_{hs}(2^-,m^2) \cdot J(q)\ ,
\ee
where $P_{hs}(2^-,m^2) := P_{11}(2^-)\Big|_{q^2=-m^2}$.

\subsubsection*{Massless spin $2^-$ }

We can set $m^2$ in \eq{a2m}, and this straightforwardly leads to the following saturated propagator
\begin{align}
\Pi &= \frac12 \int dq\, J\cdot \frac{1}{q^2}\left[ P_{11}(2^-)-P_{22}(1^-)\right] \cdot J\ .
\end{align}
The sum of the spin projectors contains nonlocal terms, i.e. terms with inverse powers of $q^2$. However, we must recall that the model has gauge symmetries, that are spelled out in detail in eqs. (D.58-60) in \cite{Baldazzi:2021kaf}.
These imply the source constraints $\partial_\rho\tau^{\alpha\rho\beta}=\partial_\rho\tau^{\rho\alpha\beta}$ and $\partial_\rho\tau^{\alpha\rho\beta}=\partial_\rho\tau^{\beta\rho\alpha}$. When these constraints are used, all the nonlocal terms disappear and the saturated propagator is
\begin{align}
\Pi &= \frac12 \int dq\, J\cdot \frac{1}{q^2}\left[ P_{11}(2^-,\eta)-P_{22}(1^-,\eta)\right] \cdot J\ .
\end{align}

\section{Various approaches to massive and massless Spin $3^-$}

\subsubsection*{Massive spin $3^-$ }

We begin with Zinoviev's the gauge invariant formulation \cite{Zinoviev83,Zinoviev:2008ck} of the Singh-Hagen Lagrangian \cite{Singh:1974qz} for massive spin 3 particle. The gauge invariant formulation is achieved by employing the fields 
\be
\{ \phi_{\m\n\rho}, \ S_{\m\n},\ A_\mu, \ \varphi \} \ ,
\ee
where $\phi_{\m\n\rho}$ is totally symmetric and $S_{\m\n}$ is symmetric. Their gauge transformations are given by
\bea
\delta\phi_{\m\n\rho} &=& \partial_{(\m} \xi_{\n\rho)} +\frac{m}{5} g_{(\m\n} \xi_{\rho)}\ ,
\nn\w2
\delta S_{\m\n} &=& m \xi_{\m\n}+\frac45 \partial_{(\m} \xi_{\n)} + m g_{\m\n} \xi\ ,
\nn\w2
\delta A_\m &=& m\xi_\m + \partial_\m \xi\ ,
\nn\w2
\delta \varphi &=& m\xi\ ,
\label{zin}
\eea
where $\xi_{\m\n}, \xi_\m, \xi$ are independent local gauge transformation parameters and $g^{\mu\nu} \xi_{\m\n}=0$.
The gauge invariant Lagrangian for massive spin 3 field is given by
\bea
\cL_{\rm Z} &=& \cL_{\rm F} + \frac16 (\partial_\m S_{\n\rho})^2 -\frac13 (\partial S)_\m (\partial S)^\m +(\partial S)_\m \partial^\m S -\frac16 (\partial_\m S)^2
\nn\w2
&& + m \big[ -\phi^{\m\n\rho} \partial_\m S_{\n\rho} +2\phi^\m (\partial S)_\m -\frac12 \phi^\m \partial_\m S \big]
+m^2 \big[ \frac12 \phi_{\m\n\rho}\phi^{\mu\n\rho} -\frac32 \phi_\m \phi^\m +\frac14 S^2\big] 
\nn\w2
&& -\frac{2}{15} \big[ (\partial_\m A_\n)^2 -(\partial A)^2\big] +m^2 \big[\phi^\m A_\m -\frac35  A_\m^2\big] +\frac23 \big[ S^{\m\n} \partial_\m A_\n -S(\partial A)\big]
\nn\w2
&& +\frac25 (\partial_\m\varphi)^2 +\frac{4m}{5}\varphi (\partial A) + m^2 (-2\varphi S+4 \varphi^2)\ ,
\eea
where $ S:= S_{\m\m}\ ,(\partial A) := \partial_\m A_\m$ and $\cL_{\rm F}$ 
is the  Fronsdal quadratic action that describes a massless spin 3 field given by
\be
\cL_{\rm F} = -\frac12 (\partial_\m \phi_{\n\rho\sigma})^2 +\frac32 (\partial\phi)_{\m\n}^2 -3(\partial\phi)_{\m\n} \partial^\m \phi^\n +\frac32 (\partial_\m\phi_\n)^2 +\frac34 (\partial\phi)^2\ ,
\label{fronsdal}
\ee
with
\be 
(\partial\phi)_{\m\n} := \partial_\lambda \phi_{\lambda\m\n}\ , \quad \phi_\mu := \phi_{\lambda\lambda\m}\ ,\quad (\partial\phi) := \partial_\m \phi_\m \ .
\ee
The Lagrangian $\cL_{\rm F}$ is invariant under the massless spin 3 gauge transformation
$\delta\phi_{abc}=\partial_{(a}\epsilon_{bc)}$ with $\epsilon^a{}_a=0$.
Imposing the gauge conditions
\be
S_{\m\n}=\frac14 g_{\m\n} S\ ,\qquad A_\m=0\ ,\qquad \varphi =0\ ,
\ee
the Lagrangian ${\cal L}$ given in \eq{zin} yields the Singh-Hagen Lagrangian for massive spin $3^-$ field,\footnote{ This Lagrangian can be expressed in different ways by using different set of auxiliary fields, on-shell they are all equivalent. The one reviewed here is the simplest version, which can be referred to as the minimal Sing-Hagen Lagrangian.}
\be
\cL_{\rm SH} = \cL_{\rm F} -\frac12 m^2 (\phi_{\m\n\rho}^2-3 \phi_\m^2) +(\partial_\m\chi)^2 + 4 m^2 \chi^2 - m\phi_\m \partial_\m \chi\ ,
\label{SH1}
\ee
where $\chi =-S/4$, which we refer to as the minimal Singh-Hagen Lagrangian for massive spin 3 field. Solving for $\chi$ from its equation of motion  and substituting back into the action yields the term $- \frac14 m^2(\partial_\mu\phi^\mu) (-\Box+4m^2)^{-1}(\partial_\nu\phi^\nu)$. Fourier transforming and expressing the result in terms of the spin projection operators gives \cite{Mendonca:2019gco}
\bea
S &=&-\frac12 \int dq\, \Big\{ \phi \cdot \Big[(q^2+m^2)P(3^-) +m^2 P(2^+)_{11} -4(q^2+m^2) P(1^-)_{11} -{\sqrt 5}m^2 P(1^-)_{14}
\nn\w2
&&-{\sqrt 5}m^2 P(1^-)_{41} - \left(\frac92 q^2+ 2m^2   -\frac{m^2 q^2}{2(q^2+4m^2)}\right) P(0^+)_{11}
\nn\w2
&& - \left(\frac12 q^2+ 2m^2   -\frac{ m^2 q^2}{2(q^2+4m^2)}\right)P(0^+)_{44}
-3 \left( (\frac{q^2}{2}+m^2)-\frac{m^2q^2}{6(q^2+4m^2)}\right) P(0^+)_{14}
\nn\w2
&& -3 \left( (\frac{q^2}{2}+m^2)-\frac{m^2q^2}{6(q^2+4m^2)}\right) P(0^+)_{41} \Big] \cdot \phi + J\cdot \phi\Big\}\ ,
\label{SHA}
\eea
giving the spin projector coefficient matrices
\bea
&& a(3^-)=-q^2-m^2
\nn\w2
&& a(2^+)= -m^2 
\nn\w2
&& a(1^-) = 
\begin{blockarray}{cc}
\begin{block}{(cc)}
4(q^2+m^2)  &  \sqrt5 m^2 \\
&\\
\sqrt5 m^2  &  0 \\
\end{block}
\end{blockarray}
\nn
\\
&& a(0^+) = 
\begin{blockarray}{cc}
\begin{block}{(cc)}
\frac12\Big[9q^2+ 4m^2 -\frac{m^2 q^2}{(q^2+4m^2)}\Big] & \frac32\Big[ q^2+2m^2-\frac{m^2q^2}{3(q^2+4m^2)}\Big]   \\
&\\
\frac32\Big[ q^2+2m^2-\frac{m^2q^2}{3(q^2+4m^2)}\Big] &  \frac12 \Big[ q^2+ 4m^2 -\frac{ m^2 q^2}{(q^2+4m^2)} \Big]\\
\end{block}
\end{blockarray}
\label{nograv2}
\eea
The nonlocal terms coming from the elimination of the auxiliary field.
Note in particular that $\det a(1^{-})=-5 m^4$ and
\bea
\det a(0^{+}) &=&\frac{20m^6}{q^2+4m^2}\ ,
\nn\w2
a^{-1}(0^+) &=&\frac{1}{40 m^6}
\begin{blockarray}{cc}
\begin{block}{(cc)}
-\big[q^4+ 7q^2 m^2 +16m^4\big]& (q^2+3m^2)(3q^2+8m^2)  \\
&\\
(q^2+3m^2)(3q^2+8m^2) &  -\big[ (9 q^4 + 39m^2 q^2+16 m^4\big]\\
\end{block}
\end{blockarray}
\label{nograv3}
\eea
From \eq{SHA} we obtain the saturated propagator
\bea
\Pi &=& \int dq\, J \cdot \Big[ \frac{1}{q^2+m^2} P(3^-) + \frac{1}{m^2} P(2^+)_{11} +\frac{4}{5m^4} (q^2+m^2)  P(1^-)_{44}
\nn\w2
&& -\frac{1}{{\sqrt 5} m^2} P(1^-)_{14} -\frac{1}{{\sqrt 5} m^2} P(1^-)_{41} +\frac{1}{40m^6} \left( (q^2)^2 +7 m^2 q^2+16m^4\right) P(0^+)_{11}
\nn\w2
&& +\frac{1}{40m^6} \left(9 (q^2)^2 +39 m^2 q^2+16 m^4\right) P(0^+)_{22}
\nn\w2
&& -\frac{1}{40m^6} (q^2+3m^2)(3q^2+8 m^2)\left( P(0^+)_{14}+P(0^+)_{41} \right) \Big] \cdot J\ .
\eea
Substituting the expressions for the spin projection operators gives (see, for example, \cite{Huang:2004we})
\bea
\Pi &=& \frac12\int dq\, J \cdot \frac{P(3^-,m^2)}{q^2+m^2} \cdot J
\nn\\
&=& \frac12\int dq\, \frac{1}{q^2+m^2}\,J^{\mu\nu\rho} \left( P_{\mu\lambda}P_{\nu\tau}P_{\rho\sigma} -\frac35 P_{\mu\nu} P_{\lambda\tau} P_{\rho\sigma} \right) J^{\lambda\tau\sigma}\ .
\label{ms3}
\eea

\subsubsection*{Massless spin $3^-$ }

The massless spin 3 field is described by the Fronsdal Lagrangian \eq{fronsdal}.  Fourier transformed action in terms of spin projection operators is 
\be
S_{\rm F} =  \frac12 \int  \Big\{ q^2 \Phi\cdot
\left[-P(3^-)+4P(1^-)_{11}+\frac92 P(0^+)_{11}+\frac32 P(0^+)_{41}+\frac32 P(0^+)_{14}+\frac12 P(0^+)_{44} \right]\cdot \Phi + J\cdot \Phi \Big\} \ .
\label{Fronsdal2}
\ee
Recalling that the totally symmetric 3rd rank tensors consist of the spin $3^1, 2_1^+, 1_1^-, 1_4^-, 014^+, 0_4+$ sectors, we see that this action has the following spin 3 gauge symmetry 
\be
\delta \Phi = P(2^+)_{11} \xi_1 
+ P(1^-)_{41} \xi_2 + P(1^-)_{44} \xi_3 + \Big[P(0^+)_{11}-3P(0^+)_{41} \Big] \xi_4\ ,
\ee
equivalent to $\delta\Phi_{cab}=\partial_{(c} \xi_{ab)}$ with $\eta^{ab}\xi_{ab}=0$, and the source constraints
\be
P(2^+)_{11}\cdot J=0\ ,\quad P(1^-)_{44} \cdot J=0\ ,\qquad P(1^-)_{14}\cdot J=0\ , \quad \left[ P(0^+)_{11}-3P(0^+)_{14}\right]\cdot J= 0\ ,
\ee
equivalent to $q^\mu J_{\mu\nu\rho} -{\rm trace}=0$. 
Using a gauge in which $P(0^+)_{11}\cdot \Phi=0$ and $P(0^+)_{41} \cdot \Phi=0$, the saturated propagator takes the form
\be
\Pi = \int dq\,  \frac{1}{q^2}\, J\cdot \Big[ -P(3^-,q) +\frac14 P(1^-,q)_{11} + 2 P(0^+,q)_{44} \Big] \cdot J
\ee 
Substituting the expressions for the spin projectors and using the source constraints, one finds
\bea
\Pi &=& \int dq\,  \frac{1}{q^2}\, J \cdot \Big[ -P(3^-,\eta) +\frac14 P(1^-,\eta)_{11} \Big] \cdot J
\nn\\
&=& \int dq\,  \frac{1}{q^2}\, J^{\mu\nu\rho}(-q) \Big[ -\eta_{\mu\lambda} \eta_{\nu\tau} \eta_{\rho\sigma} +\frac34 \eta_{\mu\nu}\eta_{\lambda\tau} \eta_{\rho^\sigma} \Big] J^{\lambda\tau\sigma}(q)\ .
\eea

\subsubsection*{Spin 3 in a gauge theory of gravity}
\label{bbh}

In section 4 we saw that the action \eq{tshs}, with parameters taken as in \eq{bbar} and all mass parameters set to zero describe only a massless spin $3^-$ particle. Here we shall compare this result with that of Baekler, Boulanger and Hehl \cite{Baekler:2006vw} where massless spin $3^-$ particle is described in the framework of a gauge theory of gravity with action quadratic in curvature, torsion and nonmetricity. Since the interest here is in massless spin 3, one keeps only the terms quadratic in curvature, that have dimensionless coefficients $c_i$, $i=\,\ldots,16$. The terms quadratic in torsion and nonmetricity, that have coefficients with dimension of square mass, can be set to zero. Then the general action takes the form 
\begin{align}
S(g,A) &=-\frac12\int d^4x\ \sqrt{|g|}\,\Big[ -a_0 F+ F^{\mu\nu\rho\sigma} \big( c_1 F_{\mu\nu\rho\sigma} 
+ c_2 F_{\mu\nu\sigma\rho} 
+ c_3 F_{\rho\sigma\mu\nu} 
+ c_4 F_{\mu\rho\nu\sigma} 
\nn\w2
&  
+ c_5 F_{\mu\sigma\nu\rho} 
+ c_6 F_{\mu\sigma\rho\nu} \big)
+ F^{(13)\mu\nu} \big(c_7 F^{(13)}_{\mu\nu} + c_8 F^{(13)}_{\nu\mu} \big)
+ F^{(14)\mu\nu} \big( c_9 F^{(14)}_{\mu\nu} 
+ c_{10} F^{(14)}_{\nu\mu}\big) 
\nn\w2
&
+ F^{(14)\mu\nu}\big(c_{11} F^{(13)}_{\mu\nu}
+ c_{12} F^{(13)}_{\nu\mu} \big)
+F^{\mu\nu}\big(c_{13} F_{\mu\nu}
+ c_{14} F^{(13)}_{\mu\nu}
+ c_{15} F^{(14)}_{\mu\nu}\big)
+c_{16}F^2
\Big]\ ,
\label{action}
\end{align}
where
\bea
F_{\mu\nu} &:=& F_{\mu\nu\lambda}{}^\lambda\ ,\quad F_{\mu\nu}^{(14)} := F_{\lambda\mu\nu}{}^\lambda\ ,\quad F_{\mu\nu}^{(13)} := F_{\lambda\mu}{}^\lambda{}_\nu\ ,\quad
F := F_{\mu\nu}{}^{\mu\nu}\ .
\eea
This theory is presented in what we called the Cartan form. When converted to the Einstein form by means of the post-Riemannian decomposition), this action becomes a quadratic expression in $R_{\mu\nu\rho\sigma}$ (the Riemann tensor), $\nabla_\lambda T_\mu{}^\rho{}_\nu$, $\nabla_\lambda Q_{\rho\mu\nu}$, plus terms cubic and quartic in torsion and nonmetricity, that we can ignore, insofar as we are only interested in propagators.
The map between the coefficients in the two forms of the action has been given in full generality in Appendix A.6 in \cite{Baldazzi:2021kaf}.
When this map is applied to the special choice of coefficients
\bea
&& c_1=c_2=\frac16\ ,\quad c_3=0\ ,\quad c_4=c_6=-c_8=-c_{10}=\frac13\ ,\quad c_5=-c_7=-c_9=-c_{12}=\frac23\ ,
\nn\\
&& c_{11}=-\frac43\ ,\quad c_{14}=-\frac14\ , \quad c_{14}=c_{15}=c_{16}=0\ .
\eea
one finds that the resulting action, at quadratic level, does not contain torsion, nor the Riemann (or Ricci) curvatures: it is a quadratic form in  $\nabla_\lambda Q_{\rho\mu\nu}$.

When linearized, it is of the form (\ref{tshs}), with coefficients

\begin{align}
\begin{tabular}{cccccccccccccccc}
$\bb_1$ & $\bb_2$ & $\bb_3$ & $\bb_4$ &  $\bb_5$ & $\bb_6$ & $\bb_7$ & $\bb_8$ &  $\bb_9$ & $\bb_{10}$ & $\bb_{11}$ &  $\bb_{12}$ & $\bb_{13}$ & $\bb_{14}$ & $\bb_{15}$ & $\bb_{16}$ \\
$\frac16$ & $\frac13$ & $-\frac23$ & $-\frac{1}{24}$ & $0$ & $-\frac16$ & $-\frac13$ & $-\frac13$ & $-\frac23$ & $\frac43$ & $0$ & $\frac23$ & $0$ & $\frac{1}{24}$ & $-\frac13$ &
\begin{minipage}[c][7mm][t]{0.1mm}%
\end{minipage}
$0$\\
\end{tabular}
\label{bbar2}
\end{align}
and $m_i=0$ for $i=1,2,3,4,5$. 
It is important to stress that the absence of torsion terms is only a property of the quadratic part of the action: the linearized theory has a gauge invariance consisting of arbitrary shifts of torsion, but this is an accidental symmetry.
\smallskip

Substituting these coefficients into \eq{tshs} gives
\begin{align}
\cL_{\rm BBH} &= -\frac{1}{12}\partial_\lambda Q_{\rho\mu\nu}\partial^\lambda Q^{\rho\mu\nu}
-\frac16\partial_\lambda Q_{\rho\mu\nu}\partial^\lambda Q^{\mu\rho\nu}
+\frac13\partial_\lambda Q_{\rho\mu}{}^\mu\partial^\lambda Q^{\rho\nu}{}_\nu
+\frac{1}{48}\partial_\lambda Q_{\rho\mu}{}^\mu\partial^\lambda Q^{\rho\nu}{}_\nu
\nn\\
&+\frac{1}{12}\partial^\lambda Q_{\lambda\mu\nu}\partial_\rho Q^{\rho\mu\nu}
+\frac16\partial^\lambda Q_{\mu\lambda\nu}\partial_\rho Q^{\mu\rho\nu}
+\frac16\partial^\lambda Q_{\mu\lambda\nu}\partial_\rho Q^{\nu\rho\mu}
+\frac13\partial^\lambda Q_{\lambda\mu\nu}\partial_\rho Q^{\mu\rho\nu}
\nn\\
& 
-\frac23 \partial_\lambda Q^{\mu\lambda\nu}\partial_\mu Q^\tau{}_{\tau\nu}
-\frac13 \partial_\lambda Q^{\mu\lambda\nu}
\partial_\nu Q^\tau{}_{\tau\mu}
-\frac{1}{48}\partial^\lambda Q_{\lambda\mu}{}^\mu\partial_\rho Q^{\rho\nu}{}_\nu
+\frac16\partial_\lambda Q_\rho{}^{\rho\lambda}\partial_\tau Q_\sigma{}^{\sigma\tau}
\ .
\label{Fr2}
\end{align}

It can be checked directly that this Lagrangian is invariant under
\bea
\delta_v Q_{\lambda\mu\nu}&=&
v_\lambda\eta_{\mu\nu}
\label{deltav}
\\
\delta_\xi Q_{\lambda\mu\nu}&=&
\xi_{\lambda\mu\nu}
\label{deltaxi}
\\
\delta_\Lambda Q_{\lambda\mu\nu}&=& 3\partial_{(\lambda}\Lambda_{\mu\nu)}
+\partial^\rho\Lambda_{\rho(\lambda}\eta_{\mu\nu)}
\label{deltaLambda}
\eea
where $\xi_{(\lambda\mu\nu)}=0$, $\xi_{\lambda\mu}{}^\mu=0$,
$\Lambda_\mu{}^\mu=0$. The relation of this Lagrangian and its symmetries to the Fronsdal Lagrangian and its symmetries can be understood as follows.
Let
\be
Q'_{\rho\mu\nu}=Q_{\rho\mu\nu} -\frac14 Q_\rho \eta_{\mu\nu}\ .
\label{redef2}
\ee
where $Q_\rho=Q_{\rho\mu}{}^\mu$ as usual.
We observe that $Q'_\rho=0$, so that this linear transformation is not invertible.
Then, one can check that
\be
\cL_{\rm BBH}(Q)=\cL_{\rm EF}(Q')=\cL_{\rm F}(S')\ ,
\ee
where in the last step we used \eq{diana}.
The noninvertibility of \eq{redef2} is the origin of the gauge symmetry \eq{deltav}.
The remaining two gauge symmetries come from the gauge symmetries of $\cL_{\rm EF}$.
Since \eq{redef2} is not invertible, we cannot deduce the transformation of $Q$ from the transformation of $Q'$, but we can do the reverse and check that the transformations \eq{deltaxi}, \eq{deltaLambda} correspond to the shift transformations \eq{hss} and tensor gauge transformations \eq{hsgtEF}. To this end, from \eq{deltaxi} and \eq{redef2}, taking into account
the tracelessness of the parameter, we find
\be
\delta_\xi Q'_{\rho\mu\nu}=\delta_\xi Q_{\rho\mu\nu}=\xi_{\rho\mu\nu}\ .
\ee
Similarly from \eq{deltaLambda} we find that
$\delta_\Lambda Q_\rho=4\partial^\mu\Lambda_{\mu\rho}$, and so inserting in \eq{redef2} gives
\be
\delta_\Lambda Q'_{\rho\mu\nu}=3\partial_{(\rho}\Lambda_{\mu\nu)}+\partial^\sigma\Lambda_{\sigma(\rho}\eta_{\mu\nu)}-\partial^\sigma\Lambda_{\sigma\rho}\eta_{\mu\nu}\ .
\ee
Now recalling that only the totally symmetric part $S$ of $Q$ is physical, we see that $\delta_\Lambda S'_{\rho\mu\nu}$ is indeed the usual higher spin symmetry.

To determine the spectrum of states described by ${\cal L}_{\rm BBH}$, we compute the 
kinetic coefficients for which we find (see \eq{euridice} for the spin projectors associated with the rows and columns)
\bea
a(3^-) &=& -2 q^2 \ ,
\nn\w4
a(0^+) &=& -2q^2 \begin{pmatrix} 
-2 & \sqrt 2 & 0\\ 
 \sqrt 2  & -1&0 \\  
 0 & 0 &0 
\end{pmatrix} 
\nn\w4
a(1^-) &=& \frac{1}{18} q^2 \begin{pmatrix}
49 & 14\sqrt 5 &  -7\sqrt 5 & 7 \sqrt{10} \\
14\sqrt 5 & 20 &  -10 & 10 \sqrt 2  \\
-7\sqrt 5 & -10 & 5 & -5\sqrt 2 \\ 
7\sqrt{10} & 10\sqrt 2 & -5 \sqrt 2 & 10 \\
\end{pmatrix}
\eea
The matrices in the spin $0^+$ and spin $1^-$ have rank 1. This gives the following gauge symmetries:
\begin{align}
\delta Q &= \sum_{k=1,2,4} \Big\{ P_{4k}(0^+)\cdot \tau_k +\Big[P_{1k}(0^+)+\sqrt2 P_{2k}(0^+)\Big]\cdot \xi_k \Big\}
\nn\\
& + \sum_{k=1,2,4,5}\Big\{ \Big[2\sqrt5 P_{1k}(1^-)-7 P_{2k}(1^-)\Big]\cdot \zeta_k^{(1)} 
+\Big[\sqrt5 P_{1k}(1^-)+7 P_{4k}(1^-)\Big]\cdot \zeta_k^{(2)} 
\nn\\
&+\Big[\sqrt{10} P_{1k}(1^-)-7 P_{5k}(1^-)\Big]\cdot \zeta_k^{(3)} \Big\}
+ P_{11}(2^{-})\cdot \alpha + P_{11}(1^+)\cdot \beta 
\nn\\
& + \sum_{k=1}^{2} \Big[ P_{1k}(2^+)\cdot \eta_k +P_{2k}(2^+)\cdot \lambda_k\Big]\ ,
\label{gt2}
\end{align}
where $\tau,\xi, \zeta,\alpha,\beta, \eta, \lambda$ are arbitrary local parameters.
We remark that the sums over $k$ are redundant: each term of those sums gives rise to the same gauge symmetries. One can thus fix an arbitrary value for $k$ in each sum.
These correspond to the gauge symmetries in (\ref{deltav},\ref{deltaxi},\ref{deltaLambda}). As a check we count 4 parameters for the $v$-transformations, 16 for the $\xi$ transformations and 9 for the $\Lambda$
transformations, totalling 29.
Since we have 3 null eigenvectors with spin 2 (2 with parity + and one with parity -)
4 with spin 1 (3 with parity - and one with parity +) and 2 with spin 0+,
and since every symmetry in a spin $J$ sector amounts to $2J+1$ parameters,
the total number of parameters in (\ref{gt2}) is also $3\times5+4\times3+2\times1=29$ parameters.

As a final remark, we note that restricting $\cL_{\rm BHH}$ to totally symmetric field $S_{\mu\nu\tau}$ (which can be achieved by gauging away the hook symmetric part $H_{\mu\nu\tau}$) we remain with
\be
{\cal L}_{\rm BHH}(S) = -\frac12 \int \Big[ \left(\partial_a S_{bcd}\right)^2 
-3\left(\partial_c S^{cab}\right)^2 +4\left(\partial_a S_b\right) \partial_c S^{cab}+\frac{17}{12} \left(\partial_a S_b\right)^2-\frac{7}{12} \left(\partial_a S^a\right)^2\Big] ,
\ee
where $S_a := \eta^{bc} S_{abc}$. Performing the field redefinition 
\be
S'_{cab}=S_{cab}-\frac14\eta_{(ca} S_{b)d}{}^d\ .
\label{redef3}
\ee
one has $\cL_{\rm BBH}(S)=\cL_{\rm F}(S')$.
Unlike \eq{redef2}, the transformation \eq{redef3} is invertible:
\be
S_{cab}=S'_{cab}+\frac12\eta_{(ca} S'_{b)d}{}^d\ .
\label{redef3}
\ee


\section{Matrix Coefficients}
\label{app:coefmat}

Recall from Table~\ref{t2} that a symmetric 3-tensor contains the following degrees of freedom:
$3^-$, $2^+_{1,2}$, $2^-_1$, $1^+_1$, $1^-_{1,2,4,5}$ and $0^+_{1,2,4}$.
We list below the kinetic coefficients describing the propagation and mixing of the
corresponding particles.

\small 
    \begin{equation}
        \begin{aligned}
    a(3^{-}) = \, -(b_{1} + b_{2})\, q^2 - ( m_{1} + m_{2})
\end{aligned}
\end{equation}

\begin{equation}
    \begin{aligned}
    a(2^{-})_{11} =& \, -\frac{1}{2}(2 b_1-b_2)\, q^2 -\frac{1}{2}(2 m_1 - m_2)
\end{aligned}
\end{equation}

\begin{equation}
    \begin{aligned}
    a(1^{+})_{11} =& \, -\frac{1}{2}(2 b_1-b_2+b_7-b_8)\, q^2 -\frac{1}{2}(2 m_1 - m_2)
\end{aligned}
\end{equation}

\begin{equation}
    \begin{aligned}
    a(2^{+})_{11} =& \,  -\frac{1}{3}(3 b_{1} + 3 b_{2} + b_{6}+b_{7}+b_{8}+b_{9})\, q^2 - (m_{1}+ m_{2})
\\
    a(2^{+})_{12} =& \, -\frac{1}{6\sqrt{2}}(4 b_{6} -2 b_{7} -2 b_{8} + b_{9})\, q^2
\\
    a(2^{+})_{22} =& \, -\frac{1}{6}(6 b_{1}-3 b_{2} +4 b_{6} +b_{7}+b_{8}-2b_{9})\, q^2 -\frac{1}{2}(2 m_{1} - m_{2})
\end{aligned}
\end{equation}

\begin{equation}
    \begin{aligned}
    a(1^{-})_{11} =& \, -\frac{1}{3}(3 b_1 +3 b_2+5b_3 + 5 b_4+5b_5)\, q^2 
-\frac{1}{3}(3 m_1 +3 m_2 + 5 m_3 + 5 m_4 + 5 m_5)
\\
    a(1^{-})_{12} =& \, -\frac{\sqrt{5}}{6}(2 b_3 - 4 b_4 - b_5)\, q^2 -\frac{\sqrt{5}}{6}(-4 m_3 + 2 m_4 - m_5)
\\
    a(1^{-})_{14} =& \, -\frac{\sqrt{5}}{6}(2b_3 + 2 b_4 + 2 b_5 + b_{10}+b_{11}+b_{12}+b_{13})\, q^2
    -\frac{\sqrt{5}}{3}(m_3 + m_4 + m_5)
\\
    a(1^{-})_{15} =& \, -\frac{\sqrt{5}}{6\sqrt{2}} (2b_3-4 b_4- b_5+b_{10}+b_{11}-2b_{12}-2b_{13})\, q^2
    -\frac{\sqrt{5}}{6 \sqrt{2}}(-4 m_3 +2 m_4 - m_5)
\\
    a(1^{-})_{22} =& \, -\frac{1}{6}(6b_1-3 b_2 +2b_3 + 8 b_4 - 4 b_5) \, q^2 
    -\frac{1}{6}(6 m_1 -3 m_2 +8 m_3 + 2 m_4 - 4 m_5)
\\
    a(1^{-})_{24} =& \, -\frac{1}{6}(2b_3 -4 b_4 - b_5 + b_{10} -2 b_{11}+ b_{12} -2 b_{13}) \, q^2 
    -\frac{1}{6}(-4 m_3 +2  m_4 -m_5)
\\
    a(1^{-})_{25} =& \, -\frac{1}{6\sqrt{2}}(2b_3 +8 b_4 -4 b_5 + b_{10} -2 b_{11}-2 b_{12} +4 b_{13}) \, q^2 
    -\frac{1}{3\sqrt{2}}(4 m_3 + m_4 - 2 m_5)
\\
        a(1^{-})_{44}  =& \, -\frac{1}{3}(3 b_1 + 3 b_2 + b_3 + b_4 + b_5 + 2 b_6 + 2 b_7
+ 2 b_8 + 2 b_9 + b_{10} + b_{11} + b_{12} + b_{13}) \, q^2 
        \\
        & \qquad
        -\frac{1}{3} (3 m_1 +3 m_2 +m_3 + m_4 + m_5)
\\
    a(1^{-})_{45} =& \, -\frac{1}{6\sqrt{2}}(2 b_3 - 4 b_4 - b_5 + 4 b_6 - 2 b_7 - 2 b_8 + b_9 + 2 b_{10} - b_{11} - b_{12} - 4 b_{13}) \, q^2 
    \\
    & \qquad
    -\frac{1}{6\sqrt{2}}(-4 m_3 +2 m_4 - m_5)
\\
    a(1^{-})_{55} =& \, -\frac{1}{6}(6 b_1 - 3 b_2 + b_3 + 4 b_4 - 2 b_5 + 2 b_6 + 5 b_7
    - 4 b_8 - b_9 + b_{10} - 2 b_{11} - 2 b_{12} + 4 b_{13}) \, q^2 
        \\
    & \qquad
    -\frac{1}{6}(6 m_1 - 3 m_2 + 4 m_3 + m_4 - 2 m_5)
\end{aligned}
\end{equation}

\begin{equation}
\begin{aligned}
    a(0^{+})_{11} =& \, -\frac{1}{3}(3 b_1 + 3 b_2 + 3 b_3 + 3 b_4 + 3 b_5 + b_6 + b_7 
    + b_8 + b_9 + 3 b_{14} + 3 b_{15} + 3 b_{16}) \, q^2 
    \\
    & \qquad
    -\frac{1}{3}(3 m_1 + 3 m_2 + 3 m_3 + 3 m_4 + 3 m_5)
\\
        a(0^{+})_{12} =& \, -\frac{1}{6\sqrt{2}}(-6 b_3 + 12 b_4 + 3 b_5 + 4 b_6 - 2 b_7 - 2 b_8 + b_9 + 12 b_{14} - 6 b_{15} + 3 b_{16}) \, q^2 
        \\
        & \qquad
        -\frac{1}{6\sqrt{2}}(12 m_3 - 6 m_4 + 3 m_5)
\\
            a(0^{+})_{14} =& \, -\frac{1}{2}(2 b_3 + 2 b_4 + 2 b_5 + b_{10} + b_{11} + b_{12} + b_{13} + 2 b_{14} + 2 b_{15} + 2 b_{16}) \, q^2 
            \\
            & \qquad
            -(m_3 + m_4 + m_5)
\\
            a(0^{+})_{22} =& \, -\frac{1}{6}(6 b_1 - 3 b_2 + 3 b_3 + 12 b_4 - 6 b_5 + 4 b_6 + b_7+ b_8 - 2 b_9 + 12 b_{14} + 3 b_{15} - 6 b_{16}) \, q^2 
            \\
            & \qquad
            -\frac{1}{6}(6 m_1 - 3 m_2 + 12 m_3 + 3 m_4 - 6 m_5)
\\
            a(0^{+})_{24} =& \, -\frac{1}{2\sqrt{2}}(-2 b_3 + 4 b_4 + b_5 - b_{10} + 2 b_{11} - b_{12}+ 2 b_{13} + 4 b_{14} - 2 b_{15} + b_{16}) \, q^2 
            \\
            & \qquad
            -\frac{1}{2\sqrt{2}}(4 m_3 - 2 m_4 + m_5)
\\
            a(0^{+})_{44} =& \, -(b_1 + b_2 + b_3 + b_4 + b_5 + b_6 + b_7 + b_8 
            + b_9 + b_{10}+ b_{11} + b_{12} + b_{13} + b_{14} + b_{15} + b_{16}) \, q^2
            \\
            & \qquad
            - (m_1 + m_2 + m_3 + m_4 + m_5)
\end{aligned}
\end{equation}

\end{appendix}


\newpage

\providecommand{\href}[2]{#2}\begingroup\raggedright\endgroup


\end{document}